\documentclass[nofootinbib]{PoS}

\usepackage{amssymb}
\usepackage{amsmath}
\usepackage{graphicx}

\newcommand{\dd}{\mathrm{d}} 
\newcommand{\Mpl}{M_\mathrm{pl}} 
\newcommand{\mpl}{m_\mathrm{pl}} 

\newcommand{\rr}{\mathrm}
\newcommand{\ns}{n_{\rr s} }
\newcommand{\As}{A_{\rr s} }

\newcommand{\fNL}{f_{\mathrm{NL}}}

\newcommand{\kp}{k_{\rr{p}}}

\newcommand{\be}{\begin{equation}}
\newcommand{\ee}{\end{equation}}
\newcommand{\ba}{\begin{align}}
\newcommand{\ea}{\end{align}}

\title{An introduction to inflation after Planck: \\ From theory to observations}

\ShortTitle{Inflation after Planck: from theory to observations}

\author{\speaker{S\'ebastien Clesse} \\ 
 Namur Center of Complex Systems (naXys), Department of Mathematics, University of Namur, Rempart de la Vierge 8, 5000 Namur, Belgium\\
        E-mail: \email{sebastien.clesse@unamur.be}}

\abstract{
These lecture notes have been written for a short introductory course on the status of inflation after Planck and BICEP2, given at the Xth Modave School of Mathematical Physics.  The first objective is to give an overview of the theory of inflation:  motivations, homogeneous scalar field dynamics, slow-roll approximation, linear theory of cosmological perturbations, classification of single field potentials and their observable predictions.   This includes a pedagogical derivation of the primordial scalar and tensor power spectra for any effective single field potential.  The second goal is to present the most recent results of Planck and BICEP2 and to discuss their implications for inflation.  Bayesian statistical methods are introduced as a tool for model analysis and comparison.  Based on the recent work of J. Martin et al., the best inflationary models after Planck and BICEP2 are presented.   Finally a series of open questions and issues related to inflation are mentioned and briefly discussed. }

\FullConference{10th Modave Summer School in Mathematical Physics,\\
		1- 5 September 2014\\
		Modave, Belgium}

\begin{document}





\tableofcontents

\section*{Introduction}

Since the 1990's and the COBE experiment, cosmology has entered into an era of high precision.  Measurements of the anisotropies in the Cosmic Microwave Background (CMB) have become increasingly accurate.  Observations of Type Ia supernovae have permitted to put in evidence the present acceleration of the Universe expansion.   Combined with the distribution of the Large Scale Structures (LSS), these observations have lead to the establishment of the standard cosmological model.  Its main ingredients are the dark matter and the dark energy (or a cosmological constant), whose origins still have to be understood.   But the model faces other theoretical problems:  the CMB temperature is the same in apparently causally disconnected regions in the sky; the Universe is very flat and its density must have been extremely fine-tuned to the critical density in the past; most Grand Unified Theories (GUT) predict that magnetic monopoles are produced in the early Universe and they should dominate the density of the Universe.   Moreover the standard cosmological model alone does not provide a mechanism for the generation of Gaussian and nearly scale-invariant initial density fluctuations.   All these problems are naturally solved if the Universe underwent an early phase of quasi-exponentially accelerated expansion, called inflation.

In the inflationary paradigm, Gaussian and nearly scale-invariant density perturbations arise naturally from the quantum fluctuations of one (or more than one) scalar field(s) slowly evolving along its (their) potential during inflation.  Besides theoretical motivations, inflation is supported by strong observational evidences.  The amplitude $\As $ and the spectral index $\ns $ of the power spectrum of the primordial curvature perturbations have been measured with accuracy by experiments probing the Cosmic Microwave Background (CMB) temperature anisotropies and polarization, such as the Planck spacecraft~\cite{Ade:2013lta,Planck:2013kta}, the Atacama Cosmology Telescope~\cite{Das:2013zf} and the South Pole Telescope~\cite{Story:2012wx}, giving $\As = 2.196^{+0.051}_{-0.06} \times 10^{-9} $ and $\ns  = 0.9603 \pm 0.0073$ in agreement with many inflation models~\cite{Martin:2013nzq}.   A strong bound have also been established on the level of local primordial non-Gaussianity, $\fNL^{\rr{loc}} = 2.7 \pm 5.8  $ \cite{Ade:2013ydc}.   In 2015, BICEP2 has claimed the detection of primordial B-mode polarization of the CMB~\cite{Ade:2014xna}, attributed to the gravitational waves produced during inflation.  The ratio between tensor and scalar perturbations $r = 0.20^{+0.07}_{-0.05}$ favors super-Planckian excursions of the scalar field responsible for inflation, called the \textit{inflaton}, and points towards an energy scale associated to inflation close to the GUT energy.   Since then, it has been argued that galactic dust could contribute more importantly to the signal than initially expected, and net detection has been transformed in a lower bound on the tensor to scalar ratio, $r < 0.11$ (95\% C.L.).  Nevertheless this question is still puzzling and future observations are required to affirm or disclaim the discovery of primordial gravitational waves~\cite{Mortonson:2014bja,Flauger:2014qra,Ade:2014xna}. 

The simplest way to realize a phase of inflation is to assume that the Universe was filled with a scalar field slowly evolving along its potential.  Depending on the shape of the potential, the primordial scalar and tensor perturbations have different statistical properties.  Despite the accurate measurements of the scalar power spectrum amplitude and spectral index and the limits on the tensor-to-scalar ratio, many single-field models are still consistent with observations, from various high energy frameworks.   

In these notes, based on a short series of lectures given in summer 2015 at the 10th Modave school of Mathematical physics, I give an overview of the theory of inflation, going from theoretical motivations to the field dynamics, both at the background and linear perturbation levels.  A particular care is given to detail the derivation of the scalar and tensor power spectra.   Results for a simple power-law scalar field potential are presented as a worked example.  The second objective is to put in evidence and discuss the status of inflation after Planck and BICEP2.  After introducing basic notions of Bayesian statistics, I will comment the recent results of J. Martin, C. Ringeval, R. Trotta and V. Vennin~\cite{Martin:2013nzq,Martin:2014lra} on model comparison.  

The lectures were intended for young researchers working in the domain of mathematical physics, not necessary familiar with the theory of inflation.   Therefore these notes may be useful for master students and young PhD in various fields of cosmology, gravitation, theoretical and mathematical physics, wanting to acquire a general and up to date culture on the topic of inflation.  I tried to make the notes as pedagogical as possible, detailing some derivations, but basic notions of cosmology and general relativity are nevertheless obvious pre-requisites. My objective is also to give them practical tricks to derive the observational predictions for any single scalar field potential that could arise in their work, and to confront their model to the most recent observations.

After introducing the motivations for an early phase of inflation in Section~\ref{sec:motivations}, I will define the usual observable quantities in Section~\ref{sec:observables}.  In Section~\ref{sec:homogeneous}, the equations governing the homogeneous dynamics for a single scalar field model are given and the common slow-roll approximation is introduced.    The linear theory of cosmological perturbation will be used in Section~\ref{sec:perturbations} to derive the power spectrum of scalar and tensor perturbations at first order in slow-roll.  Observational predictions for a simple large field model are calculated in Section~\ref{sec:largefield} as an example.  In Section~\ref{sec:models} a classification of single-field potentials is presented.  The most important results of Planck and BICEP2 experiments are presented and discussed in Section~\ref{sec:observations}.   In Section~\ref{sec:modelanalysis}, after an introduction to Bayesian analysis for model constraints and comparison, I will comment on the favored single-field scenarios based on Planck only and on Planck+BICEP2.   Finally I will mention in Section~\ref{sec:openquest} a few open theoretical questions linked to inflation. 

\vspace{5mm}
Note that a lot of material has been borrowed directly from the first and second chapters of my PhD thesis~\cite{Clesse:2011jt}.   These notes are also based on previous but recent lectures notes and reviews by D. Langlois~\cite{2010LNP...800....1L}, A. Linde~\cite{Linde:2014nna}, D. Baumann and L. McAllister~\cite{Baumann:2014nda}, to which the interested reader may refer to get further information about the topic.  An exhaustive analysis of all single field inflationary models proposed so far can be found in Ref.~\cite{Martin:2013tda}.  As already mentioned, the section on model analysis and Bayesian theory is based on Refs.~\cite{Martin:2013nzq,Martin:2014lra}.

\vspace{5mm}
 Any comments about these lecture notes are welcome.

\newpage
\section{Motivations} \label{sec:motivations}

The standard hot Big-Bang cosmological model suffers from several problems.  In this section, they are briefly explained, and then it is shown how a sufficiently long phase of  inflation naturally solve them.  

\subsection{Problems of the standard cosmological model}
\subsubsection{Horizon problem}  \label{sec:horizon}

\begin{figure}[ht] 
	\begin{center}
	\includegraphics[height=70mm]{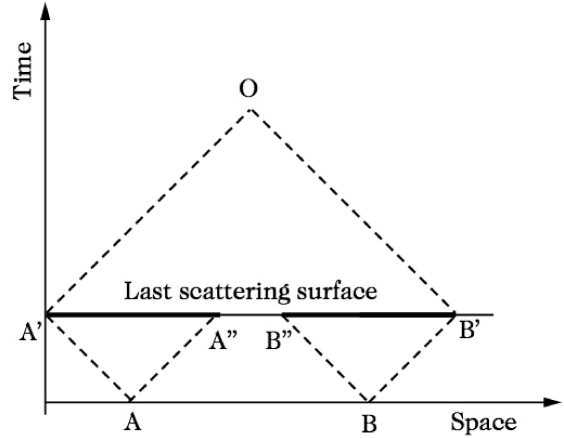}
	\caption{Scheme~\cite{PeterUzan} illustrating the horizon paradox.   The CMB is observed from the hypersurface $t=t_{\mathrm 0}$. The $AB'$ region at last scattering is isothermal in the CMB sky, although it appears to be constituted of causally disconnected patches.}
	  \label{fig:horizon}
	\end{center}
\end{figure}

Before to express and explain the horizon problem, it is useful to define the various notions of horizon in cosmology.  It is convenient to define the conformal time
\begin{equation}
\eta(t) = \int_{t_{\mathrm i}} ^{t} \frac {\dd t'}{a(t')} ,
\end{equation}
where $a$ is the scale factor and $t$ denotes the cosmic time. It is the 
comoving distance covered by light between an initial hyper-surface at time $t_{\mathrm i}$ and the hyper-surface at time $t$.   Assuming that the Universe evolution starts at $t_{\rr i}$, two points separated by a comoving distance larger than the conformal time $\eta$ do not have a causal link.   Usually, the initial hyper-surface is set at the Planck-time, and all points separated by a comoving distance larger than $\eta$ are said to be \textit{causally disconnected}.  For an observer in $O$ at a time $t_{\mathrm 0}$  (see Fig.~\ref{fig:horizon}),  $\eta(t_{\mathrm 0})$ is the comoving radius of the sphere centered in $O$ separating particles causally connected to the observer of particles causally disconnected.  $\eta(t)$  is called the \textit{comoving horizon} or the \textit{particle horizon}.   It is important to distinguish between the \textit{particle horizon} and the \textit{event horizon}, which is, for the observer, the hypersurface separating the universe in two parts, the first one containing events that have been, are or will be observable, the second part containing events that will be forever unobservable.    Mathematically, the \textit{event horizon} exists only if the integral 
 \begin{equation} \label{evenements}
 \int_{t_{\mathrm i}} ^\infty \frac{\dd t'}{a(t')} 
 \end{equation}
 converges.  Finally, it is useful to define the \textit{comoving Hubble radius}, $ 1/(aH)$. 
 It is smaller than the conformal time, that is the logarithmic integral of the Hubble radius.

The \textit{horizon problem}  is linked to the isotropy of the CMB and can be expressed in the following way:  how to explain that regions in the CMB sky have the same temperature whereas their angular separation is too large to correspond to causally connected patches at the time of last scattering, assuming that the standard cosmological model is valid down to the Planck time?

In the standard cosmological model, the early Universe is dominated by the radiation and the chemical potentials can be neglected most of the time.   One has therefore \hbox{$ a T = \mathrm{constant} $}, and in a comoving coordinate system, any physical distance grows like
\begin{equation} d (t) = \frac {T(t_{\mathrm 0})}{T(t)} d(t_{\mathrm 0})~. \end{equation}
The temperature of CMB photons today is $T_0 \approx 2.7 K \approx 2.3 \times 10^{-13} \mathrm{GeV}$. 
On the other hand, assuming that the expansion rate is governed by the standard cosmological model at every time, the radius of the observable universe, i.e. the radius of the spherical volume in principle observable today by an observer at the center of the sphere, is $d_{H_0}(t_0) \approx 10^{26} 
\mathrm{m} 
$.   At the time of last scattering $t_{\rr{LSS}}$, the radius of the observable universe was
 \begin{equation} 
 d_{H_0}(t_{\rr{LSS}}) \approx 7 \times 10^{22} \rr{m}~. 
 \end{equation}
Under the same assumption, at recombination, the maximal distance between two causally connected points would roughly be
\begin{equation} 
d_{\rr H_{\rr{LSS}}} (t_{\rr{LSS}}) \approx 2 \times 10^{21} \rr{m}~. 
\end{equation}
At last scattering, our observable Universe would therefore have been constituted of about $ 10^5$ causally disconnected regions.  But CMB photons emerging from these regions are observed to have all the same temperature, to a $10^{-5}$ accuracy.   
At the Planck time, the number of causally disconnected patches would have been much larger, about $10^{89} $.

\subsubsection{Flatness problem}

The Einstein equations in a homogeneous and isotropic FLRW Universe give the Friedmann-Lema\^itre (FL) equations 
\begin{eqnarray} \label{eq:FL}
H ^2 & = & \frac {8 \pi }{3 \mpl^2 } \rho - \frac{K}{a^2} + \frac 1 3 \Lambda~, \\  \label{eq:FL2}
\frac {\ddot a} a & = & - \frac {4   \pi }{3 \mpl^2}  (\rho + 3 P ) + \frac 1 3 \Lambda~,
\end{eqnarray}
where $H\equiv \dot a/a $ is the Hubble expansion rate, $a$ the scale factor, $K = 0,\pm 1$ is the curvature, $\rho$ the energy density and $P$ the pressure.  $\mpl$ is the 4-th dimensional Planck mass, which should not be confused with the reduced Planck mass $\Mpl \equiv \mpl / \sqrt{8\pi}$. 
Starting from the Friedmann-Lema\^itre equations and neglecting the cosmological constant\footnote{This is a good approximation because $\Lambda$ dominates the energy density only at late times.}, one can find the evolution equation for the curvature density $\Omega_{\rr K}$, defined as
\be
\Omega_{\rr K} \equiv  - \frac {K}{a^2 H^2} = 1- \Omega.
\ee
where $\Omega \equiv 8 \pi \rho / (3 H^2 \mpl^2)$.  One gets
\begin{equation}
\frac{\dd \Omega_{\rr K} } { \dd \ln a} = (3 w + 1 ) ( 1 - \Omega_{\rr K} ) \Omega_{\rr K}~,
\end{equation}
where $w \equiv P/\rho$ is the equation of state parameter.  This equation is easily integrated when $w$ is constant.  One has 
\begin{equation}
\frac{ \Omega_{\rr K 0} }{\Omega_{\rr K} (a) } = ( 1 - \Omega_{\rr K 0} ) \left( \frac{a}{a_0} \right)^{(-1 - 3 w)} + \Omega_{\rr K 0} ~,
\end{equation}
where $\Omega_{\rr K 0} $ is the curvature today.  Since it is constrained by observations ($|\Omega -1| \lesssim 0.01$~\cite{Komatsu:2010fb}) one has roughly at radiation-matter equality
\begin{equation}
|\Omega(a_{\rr{eq}}) -1|  \lesssim 3 \times 10 ^{-6}~,
\end{equation}
and at the Planck time,
\begin{equation}
|\Omega (a_{\rr{p}}) -1|  \lesssim 10^{-60}~.
\end{equation}
If the Universe is not strictly flat, the $\Lambda$CDM model does not explain why the spatial curvature is so small.  

\subsubsection{Problem of topological defects}

\begin{figure}[ht]
 	\begin{center}
	\includegraphics[height=90mm]{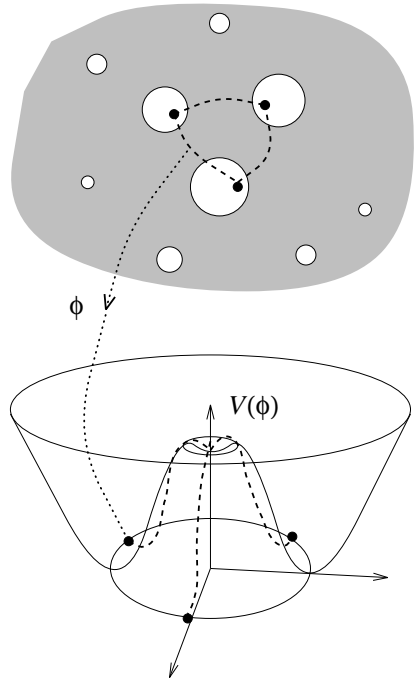}
	\caption{Illustration~\cite{Hindmarsh:1994re} of the formation of cosmic strings due to the breaking of the group $U(1)$ into $\{ \rr{id} \}$.   After the transition, the Higgs field $\phi$ takes a different value at each point in space.   When the Higgs field makes a complete loop in the field space along a closed path in the real space, there exists a point inside the path for which the phase is not defined.  At this point, the Higgs field vanished, the symmetry is restored and the resulting string configuration contains energy.  This process is called \textit{ the Kibble mechanism} (for a review, see~\cite{Hindmarsh:1994re}).}    \label{fig:kibble}
	\end{center}
\end{figure}

In Grand Unified Theories (GUT), the standard model of particle physics results from several phase transitions induced by the spontaneous breaking of symmetries.  Such a symmetry breaking is triggered during the early Universe's evolution due to its expansion and cooling, and they can lead to the formation of topological defects like domain walls, cosmic strings and monopoles.   These defects correspond to configurations localized in space for which the initial symmetry remains apparent (see Fig.~\ref{fig:kibble}).

Let us consider the symmetry breaking of a group $\mathcal G$ resulting to an invariance under the sub-group $\mathcal H$:  $ \mathcal G \rightarrow \mathcal H $.  The vacuum manifold $\mathcal M$ is isomorphic to the quotient group $ \mathcal G / \mathcal H$~\cite{Nakahara:1990th}.  \textit{Domain walls} are formed when the 0th-order homotopy group of $\mathcal M$ is not trivial.  They can be due to the breaking of a $Z_{2}$ symmetry, or if the resulting vacuum contains several distinct elements.   \textit{Cosmic strings} are formed when the first homotopy group of $\mathcal M$ is not trivial, for instance for the breaking scheme $U(1) \rightarrow \{ \rr{Id} \}$.   \textit{Monopoles} are formed when the second homotopy group $\pi_2 (\mathcal M)$ of the vacuum manifold is not trivial.  This is the case for the breaking of a $SO(3)$ symmetry into $\mathcal H = \{ \rr{Id} \} $.  For higher homotopy groups, the resulting topological defects are called \textit{textures}.  

Groups involved in GUT are such that the first and second homotopy groups are trivial, $ \pi_1 (\mathcal G) \sim \pi_2 (\mathcal G) \sim \rr{Id}$.  In the SM, there remains a U(1) invariance corresponding to electromagnetism.  The first homotopy group of U(1) is $\pi_1 \left[ U(1)\right] \sim Z $.  Therefore, by using the property of homotopy groups~\cite{PeterUzan}
\begin{equation} \pi_n(\mathcal G) \sim \pi_{n-1} (\mathcal G) \sim {Id.} \Rightarrow \pi_n (\mathcal M) \sim \pi_{n-1} (\mathcal H)~,
\end{equation}
one obtains that the second homotopy group of the vacuum manifold corresponding to the breaking of a GUT group is not trivial.  That induces necessarily the formation of monopoles~\cite{Langacker:1980kd}.

However, monopole annihilation has been found to be very slow \cite{Zeldovich:1978wj,PhysRevLett.43.1365}.  As a consequence, their energy density today should be 15 orders of magnitude larger than the current energy density of the universe.   Domain walls can also lead to catastrophic scenarios, but they can be avoided in the schemes of symmetry breaking in GUT.   Cosmic strings are observationally allowed, but their contribution to the CMB angular power spectrum~\cite{Kaiser:1984iv} is constrained~\cite{Battye:2010xz}.  

\subsubsection{Nearly scale-invariance of the primordial scalar power spectrum}

The density perturbations at the origin of the CMB temperature fluctuations start to oscillate when their size becomes smaller than the Hubble radius.   On the contrary, the perturbations whose wavelength is much larger at recombination have remained constant and thus conserve their initial amplitude.  In the CMB angular power spectrum, these super-Hubble perturbations correspond to temperature fluctuations at low multipoles  ($l \lesssim 20$).   The CMB temperature fluctuations at large angular scales therefore directly probe the initial state of those density perturbations.   

With CMB observations, it has been established that the primordial power spectrum of density perturbations is (nearly) scale invariant.   The present measurements of the shape of the primordial power spectrum will be given in details in Section~\ref{sec:observables}.   
This constrains the possible physical processes at the origin of the initial density perturbations. 


\subsubsection{Absence of iso-curvature modes}

There are two different kinds of primeval fluctuations:  the \textit{curvature} (or \textit{adiabatic}) and \textit{iso-curvature} (or \textit{entropic}).  

The adiabatic density fluctuations are characterized as fluctuations in the local value of the spatial curvature (hence the name of \textit{curvature} perturbations).  By the equivalence principle, all the species contribute to the density perturbation and one has for any fluid $f$,
\begin{equation}
\frac{\delta \rho}{\rho} = \frac{\delta n_f}{n_f} = \frac{\delta s}{ s}~,
\end{equation}
where $s \equiv S / a^3 $ is the entropy density.  Furthermore, one can write
\begin{equation}
\delta \left(  \frac{ n_f}{s} \right) = \frac{\delta n_f}{ s} - \frac{n_f \delta s} {s^2} = 0~.
\end{equation}
That means that the fluctuation in the local number density of any species relative to the entropy density vanishes. 

The entropic fluctuations are perturbations for which $\delta \rho = 0 $ and therefore they are not characterized by fluctuations in the local curvature (hence the name \textit{iso-curvature}).   They correspond to fluctuations in the equation of state.   
 
 With observations one has determined that the CMB temperature fluctuations are sourced by curvature perturbations, and one has constrained the possible contribution of iso-curvature perturbations~\cite{Komatsu:2010fb}.  The mechanism producing the initial inhomogeneities needs to generate only (or at least mostly) curvature perturbations.

\subsubsection{Why are perturbations Gaussian?}

The statistical properties of the CMB anisotropies are encoded in the power spectrum of the temperature fluctuations, i.e. in the two-point correlation function in the Fourier space, as well as in the three-point (bispectrum), four-point (trispectrum), and higher order correlation functions.   But if the fluctuations follow a Gaussian statistic, these latter ones
 are all vanishing.   CMB experiments have not detected with a high significance a non-zero value neither for the three-point neither for higher-order correlation functions.  Since 
 the temperature fluctuations in the CMB are induced by density perturbations, 
 the mechanism generating them must be to be such that their statistics is Gaussian.

\subsection{Cosmic Inflation}

Inflation is a phase of quasi-exponentially accelerated expansion of the Universe.  By combining the Friedmann-Lema\^itre equations and assuming $K=0$, one obtains a necessary condition for inflation to take place, 
\begin{equation} \label{condition} 
\ddot a > 0 \Longleftrightarrow \rho + 3 P < 0~.
\end{equation}

The amount of expansion during inflation is measured in term of the number of \textit{e-folds}, defined as
\begin{equation}
N(t) \equiv \ln \left[ \frac {a(t)}{a_{\mathrm i}} \right],
\end{equation}
where $a_{\mathrm i} $ is the scale factor at the onset of inflation.   

The inflationary paradigm is well motivated because it provides a solution to all the problems of the standard cosmological model that have been mentioned:
\begin{itemize}
\item \textbf{The horizon problem}:  Inflation solves naturally this paradox if the number of e-folds of expansion is sufficiently large.  Indeed, isothermal regions in the CMB apparently causally disconnected at recombination, can actually be in causal contact because of a primordial phase of inflation.  Assuming that the expansion was exponential during inflation,
\begin{equation} a(t) = a_{\rr i} e^{H \Delta t }~, \end{equation}
(it will be shown later that this condition is nearly satisfied) one can evaluate the number of e-folds required to solve the horizon problem.   At the end of inflation, the size of the current observable Universe $d_{H_0}$ must have been smaller than the size of a causal region at the onset of inflation $d_{H_{\rr i}} $,
\begin{equation}
d_{H_0}(t_0)  \frac{a_{\rr{end}}  }{a_0}  <  d_{H_{\rr i}} \frac{a_{\rr{end} }  } {a_{\rr i}} = d_{H_{\rr i}}(t_{\rr i}) \rr e^{N}~,
\end{equation}
where $a_{\rr{end}}$ is the scale factor at the end of inflation.  If inflation ends at the Grand Unification scale ($\rho_{\rr{end}}^{1/4} \sim 10^{16}
$ GeV), one needs
\begin{equation}
N \sim \ln \left( \frac{T_0  d_{H_0}(t_0)  }{T_{\rr{end}}   d_{H_{\rr i}}(t_{\rr i}) }  \right) \gtrsim 57~,
\end{equation}
where $T_0$ is the photon temperature today, and for which we have assumed $ d_{H_{\rr i}}(t_{\rr i}) \sim l_{\rr{Pl}} T_{\rr Pl} / T_{\rr{end}}  $, where $l_{\rr {Pl}}$ and $T_{\rr {Pl}}$ are respectively the Planck length and the Planck temperature.   If this condition is satisfied, the entire observable Universe emerges out of the same causal region 
before the onset of inflation.  

\item \textbf{The flatness problem}:  During inflation, the Universe can be extremely flattened.    Indeed, if we assume $H$ to be almost constant during inflation, one has 
\begin{equation}
| \Omega_{\rr K} (a_{\rr{end} } ) | = | \Omega_{\rr K} (a_{\rr{i} } ) | \rr e^{- 2 N}~,
\end{equation}
and with $N \gtrsim 70 $ and a curvature of the order of unity at the Planck scale, the flatness problem is  naturally solved. 

\item \textbf{Topological defects}:  During inflation, topological defects are diluted due to the volume expansion and are conveniently stretched outside the observable Universe. 
\item \textbf{The primordial power spectrum}:  Models of inflation generically predict a nearly scale invariant power spectrum of curvature perturbations, and thus can provide good initial conditions for the density fluctuations in the radiation era. We will give more details on how to calculate the curvature power spectrum in the case of single-field inflation later in the notes.
\item \textbf{Gaussian perturbations}:  In inflation, all the structures in the Universe are seeded by quantum fluctuations.  As the Universe grows exponentially, the quantum-size fluctuations become classical, are stretched outside the Hubble radius, and source the CMB temperature fluctuations.  All the pre-inflationary classical fluctuations are conveniently stretched outside the Hubble radius today and can be safely ignored.   The Gaussian statistic of the perturbations therefore takes its origin in the Gaussian nature of the quantum field fluctuations.  
\item \textbf{Iso-curvature modes}:   Most models of inflation source only curvature perturbations. Nevertheless, for some models (like multi-field models), the iso-curvature mode contribution can be potentially important and eventually observable (e.g. in Ref.~\cite{Langlois:1999dw}).  In multi-field models, these are induced by field fluctuations orthogonal to the field trajectory.
\end{itemize}

\section{Observables}  \label{sec:observables}

The shape of the CMB angular power spectrum is sensitive to the initial conditions of the density and curvature fluctuations.   
Inflation provides these initial conditions and can therefore be confronted to CMB observations.  

Observations of CMB temperature anisotropies and polarization have permitted to measure precisely the amplitude of the power spectrum of scalar curvature perturbations, its spectral index that quantifies the deviation from scale-invariance, and to put strong limits on the ratio between curvature and tensor metric perturbations.  We skip here the details 
of the CMB theory relating initial conditions to temperature and polarization anisotropies.  This is a difficult topic that goes beyond the scope of these notes.  Instead we focus directly on the quantities that are commonly used to constrain inflation.

\subsection{Power spectrum of primordial curvature perturbations} \label{sec:powerspectrum}

The primordial power spectrum of the curvature perturbation $\zeta$ is defined from,
\begin{equation}
\langle \hat \zeta(\mathbf k) \hat \zeta (\mathbf k' ) \rangle = (2 \pi)^3 P_\zeta(k) \delta^3 (\mathbf k - \mathbf k')~,
\end{equation}
where $\hat \zeta (\mathbf k)$ is the 3-dimensional Fourier transform of $\zeta (\mathbf x )$.  The spectral  index of this power spectrum $n_{\rr s}$ is defined as
\begin{equation}
n_{\mathrm s} \equiv 1+ \left. \frac {\dd \ln \left[ k^3 P_{\zeta} (k) \right] } { \dd \ln k } \right|_ {k_*} ~,
\end{equation}
where $k_*$ is a pivot scale in the observable range, $k_* = 0.05 \ \rr{Mpc}^{-1} $ for Planck.  A power spectrum increasing on large angular scales ($n_{\rr s}< 1$) is called  \textit{red-tilted}, if it increases with small scales it is called \textit{blue-tilted}. 
Deviation from a scale invariant primordial power spectrum have been detected by recent CMB experiments. The power spectrum is observed to be red-tilted, and the case $n_{\rr s } = 1$ is today ruled out.   The 1-$\sigma$ bound on the spectral index measured by WMAP was~\cite{Komatsu:2010fb} were $n_{\rr s} = 0.968 \pm 0.012 $.  Planck has improved by roughly a factor two the measurement of the spectral index~\cite{Ade:2013lta,Planck:2013kta}, giving $n_{\rr s} = 0.9603 \pm 0.0073 $. 

On the other hand, the power spectrum amplitude measured by Planck is 
\begin{equation}  
\As \equiv \mathcal P_\zeta (k_*) \equiv \frac{k_*^3}{2 \pi^2} P_\zeta(k_*) =  2.196^{+0.051}_{-0.06} \times 10^{-9} .
\end{equation}

\subsection{Tensor-to-scalar ratio}

The tensor metric perturbations, characterized by a power spectrum $P_{h}(k)$ at the end of inflation, can also affect the CMB angular power spectrum, and especially the B-mode polarization of the CMB (for details, see e.g.~\cite{PeterUzan}).
CMB observations by WMAP and Planck have established a strong limit on the amplitude of primordial gravitational waves.  Usually, this limit is given as an upper bound on the ratio $r$ between tensor and scalar power spectrum amplitudes at the pivot scale.  After WMAP, the 2-$\sigma$ bound on the tensor to scalar ratio was~\cite{Komatsu:2010fb}, 
\begin{equation} 
r \equiv \frac{P_{h}} {P_{\zeta}}  < 0.24~.
\end{equation}
With WMAP polarization data plus Planck observations of temperature anisotropies, this limit is reduced to~\cite{Ade:2013lta,Planck:2013kta} $ r < 0.11 $, whereas BICEP2 claimed in 2014 a tensor to scalar ratio $ r = 0.20^{+0.07}_{-0.05}$

\subsection{Other observables}

There are other observable quantities of interest to further constrain inflation models.   Some of  these are defined below:
\begin{itemize}
\item The running of the spectral index $\alpha_{\rr s}$:  it is defined as 
\begin{equation}
\alpha_{\rr s} \equiv \left. \frac{\dd n_{\rr s} }{\dd \ln k} \right|_{k=k_*}~.
\end{equation}
Present data are compatible with $\alpha_{\rr s} = 0$.   Planck $1\sigma$ limits give~\cite{Ade:2013lta,Planck:2013kta} $ \alpha_{\rr s} = -0.0134 \pm 0.0090$.
\item The $f_{\rr{NL}} $ parameter:   this parameter characterizes the amplitude of the so-called local form of the bispectrum of $\zeta$, 
\begin{equation}
B_\zeta = \frac 6 5 f_{\rr{NL}}  \left[ P_\zeta(k_{\rr 1}) P_\zeta(k_{\rr 2}) + (\rr{2 \ perm.}) \right]~,
\end{equation}
defined as the Fourier transform of the three-point correlation function, 
\begin{equation}
\left\langle \prod_{i=1}^3 \zeta(\mathbf{ k}_i)   \right\rangle = (2 \pi )^3 \delta^3 \left(\sum_{i=1}^3 \mathbf{ k}_i \right) B_\zeta(k_{\rr 1},k_{\rr 2},k_{\rr 3})~.
\end{equation}
A non-zero bispectrum results from non-Gaussian curvature perturbations.  Inflation can be a source of small non-Gaussianities, but also the reheating phase, eventual cosmic strings, and various astrophysical processes.  In the \textit{squeezed limit}, corresponding to $ k_{\rr 3} \ll k_{\rr 1} \simeq k_{\rr 2} $, it has been shown that all single-field models of inflation yield to $f_{\rr{NL}}^{\rr{loc}}  = \frac 5 {12} (1-n_{\rr s} ) \simeq 0.02 $~\cite{Gangui:2002qc,Creminelli:2004yq}.   For multi-field models the  $f_{\rr{NL}}^{\rr{loc}} $ value can take higher values, potentially detectable by experiments.  For the other processes, the amplitude should be $f_{\rr{NL}}^{\rr{loc}}  \sim \mathcal O (1)$ (see~\cite{Komatsu:2010hc} for a review), thus a convincing detection of $f_{\rr{NL}}^{\rr{loc}}  \gg 1$ would have ruled out most\footnote{However, non-Gaussianities in single field models could be generated by trans-planckian effects~\cite{Collins:2009pf} or slow-roll violation~\cite{PhysRevD.83.103511}} single field inflation models.   For WMAP, the best limit was obtained in Ref.~\cite{Sugiyama:2011jt} 
\begin{equation}
f_{\rr{NL}}^{\rr{loc}}   = 32 \pm 21 \ \rr{(68 \% C.L. )}~.
\end{equation}
The Planck satellite have reduced the error bars by a factor of four and the present limits are $\fNL^{\rr{loc}} = 2.7 \pm 5.8  $ \cite{Ade:2013ydc}.

\item The  $\tau_{\rr{NL}} $ parameter:   this parameter characterizes one of the amplitudes of the local-form trispectrum of $\zeta$, 
\begin{equation}
T_\zeta = \tau_{\rr{NL}}  \left[ P_\zeta(\mathbf {k}_{\rr 1} + \mathbf{ k}_{\rr 3}) P_\zeta(k_{\rr 3})  P_\zeta(k_{\rr 4})+ (\rr{11 perm.}) \right]~,
\end{equation}
which is the Fourier transform of the four-point correlation function, 
\begin{equation}
\left\langle \prod_{i=1}^4 \zeta(\mathbf{ k}_i)   \right\rangle = (2 \pi )^3 \delta^3 \left(\sum_{i=1}^4 \mathbf{ k}_i \right) T_\zeta(k_{\rr 1},k_{\rr 2},k_{\rr 3},k_{\rr 4})~.
\end{equation}
The present limits are still relatively weak ($ \tau_{\rr{NL}} < 2800$ (95\% C.L.) ) and in agreement with  $ \tau_{\rr{NL}} = 0$ 
\end{itemize}

\section{1-field inflation:  Background dynamics} \label{sec:homogeneous}



The easiest realization of the condition (\ref{condition}) is to assume that the Universe is filled with an unique homogeneous scalar field $ \phi $, called the \textit{inflaton}. The Lagrangian reads
\begin{equation}
 \mathcal L = - \sqrt {-g} \left[ \frac 1 2 \partial_{\mu} \phi \partial^{\mu} \phi   + V(\phi) \right] ~,
 \end{equation}
where $V(\phi)$ is the scalar field potential and $g$ is the determinant of the FLRW metric.  The equation of motion (e.o.m.) for this Lagrangian is the Klein-Gordon equation in an expanding spacetime,
\begin{equation} \label{KGtc}
\ddot \phi + 3 H \dot \phi + \frac {\dd V}{\dd \phi} = 0~. 
\end{equation}
On the other hand, the energy momentum tensor reads
\begin{equation}
T_{\mu \nu} = - \frac 2 {\sqrt{-g}} \frac {\delta  \mathcal L }{\delta g_{\mu \nu} }~.
\end{equation}
The energy density and the pressure are therefore
\begin{eqnarray}  \label{eq:rho_inf}
 \rho & = & \frac{\dot \phi ^2}  2 + V(\phi)~, \\  \label{eq:P_inf}
 P & = & \frac {\dot \phi ^2} 2 - V(\phi)~.
 \end{eqnarray}
The condition (\ref{condition}) is satisfied if the scalar field evolves sufficiently slowly, so that
 $\dot \phi ^2 \ll V(\phi) $.
The expansion is governed by the Friedmann-Lema\^itre equations  
\begin{equation} \label{eq:FLtc1}
H^2 = \frac {8\pi }{3 \mpl^2}  \left[ \frac 1 2 \dot
\phi^2    + V(\phi) \right] ~, 
\end{equation}
\begin{equation} \label{eq:FLtc2}
\frac{\ddot a }{a} = \frac {8\pi}{3 \mpl^2} \left[ -  \dot \phi^2
 + V(\phi ) \right]~.
\end{equation}
One can rewrite them in e-fold time $ N$
\begin{eqnarray}
H^2 & = &  \dfrac {8\pi }{ \mpl^2}  \frac{V(\phi)}{3 - \dfrac {4 \pi}{\mpl^2} \left( \dfrac{\dd \phi}{\dd N}  
\right)^2}~,\\
 \frac{1}{H}  \frac {\dd H}{\dd N} & = & - \frac{4 \pi}{\mpl^2} \left(  \frac{\dd \phi}{\dd N}\right)^2~,\\
\frac 1 {3- \frac{4 \pi}{\mpl^2} \left( \frac{\dd \phi}{\dd N} \right)^2 }\frac{\dd^2 \phi}{\dd N^2}  + \frac{\dd \phi}{\dd N} & = & -  \frac { \mpl^2} {8\pi } \frac{\dd \ln V}{\dd \phi}~,
\end{eqnarray}
and one sees that the scalar field evolves independently of the Hubble rate dynamics. 

 \subsection{Slow-roll approximation}

For inflation to be very efficient, the kinetic terms in the F.L. equations must be very small compared to the potential.   The \textit{slow-roll approximation} consists in neglecting the kinetic terms and the second time derivatives of the field, 
\begin{equation}
 \dot \phi ^2 \ll V(\phi)~, \hspace{30mm} \ddot \phi \ll 3H\dot \phi~.
 \end{equation}
In the slow-roll regime, one has therefore 
\begin{eqnarray}\label{sr1} 
H^2 & = & \frac {8 \pi }{3m_{\mathrm p}^2} V(\phi)~,\\
\label{sr2}
 3H \dot \phi & = & - \frac {\dd V}{\dd \phi}~.
\end{eqnarray}
Using the number of e-folds as a time variable, the field evolution is governed by
\begin{equation}
\frac {\dd \phi}{\dd N} = -\frac { \mpl^2}{8\pi } \frac 1 V \frac {\dd V}{\dd \phi}~.
\end{equation}
One sees that a large number of e-folds is realized in a small range of $\phi$ when the logarithm of the potential is very flat.

The slow-roll regime is an attractor~\cite{Ringeval:2005yn} such that typically a few e-folds after the onset of inflation, the slow-roll approximation is valid.  As shown later, studying inflation in the slow-roll regime is very convenient because in this case model observable predictions for the scalar and tensor power spectra are directly related to the scalar field potential and its derivatives.  Now let us introduce the Hubble-flow functions~\cite{Leach:2002ar}, 
\begin{eqnarray}
 \epsilon_{1}  & \equiv &  - \frac {\dot H}{H^2} < 1 \Longleftrightarrow \ddot a > 0 ~,\\
 \epsilon_{n+1} & \equiv & \frac {\dd \ln |\epsilon_{n}| }{\dd N}~. 
\end{eqnarray}
Using these functions,  the F.L. and K.G. equations can be rewritten as
\begin{eqnarray} H^2 & = & \frac {8 \pi}{m_{\mathrm p} ^2} \frac {V}{3-\epsilon_1}~, \\
\dot \phi & = & \dfrac {-1} { \left(3+\dfrac 1 2 \epsilon_2 - \epsilon_1 \right) H } \frac {\dd V}{\dd \phi}~, 
\end{eqnarray}
and one sees that the slow-roll regime is recovered when 
\begin{equation} \epsilon_1 \ll 3~, \hspace{2cm} \epsilon_2 \ll 6 - 2 \epsilon_1 ~. \end{equation}
One sees also that $\epsilon_1 < 3 $ is required for satisfying the condition $H^2 >0 $.  In  the slow-roll approximation, they can be expressed as a function of the potential and its derivatives.  For the first and second Hubble-flow functions, one has~\cite{Liddle:1994dx}
\begin{equation} \begin{split}
\epsilon_{\rr 1} (\phi) & \simeq  \frac{\mpl^2}{16 \pi} \left( \frac{1}{V} \frac{\dd V}{\dd \phi}  \right)^2 + \mathcal O (\epsilon_i ^2) ~,\\
\epsilon_{\rr 2} (\phi) & \simeq \frac{\mpl^2}{4 \pi} \left[\left( \frac{1}{V} \frac{\dd V}{\dd \phi}  \right)^2 - \frac{1}{V} \frac{\dd^2 V}{ \dd \phi^2}  \right] + \mathcal O (\epsilon_i ^2) ~.
\end{split} \label{eq:slowrollparams}
\end{equation}
The Hubble flow functions are usually referred as the \textit{slow-roll parameters}. 
Finally, let remark that there exists several definitions for the slow-roll parameters, e.g. $ \epsilon $ and $ \eta_{\rr{SR}}$ defined as
\begin{eqnarray} 
\epsilon & \equiv & \epsilon_1~, \\
 \eta_{\rr{SR}} & \equiv & - \frac{\ddot \phi}{H \dot \phi} = \epsilon - \frac{\dot \epsilon}{2 H \epsilon} \simeq \frac{ \mpl^2}{4 \pi} \frac{V''}{V}  ~,
 \end{eqnarray}
such that the relation $\epsilon_2 =  -  \eta_{\rr{SR}} + 4 \epsilon $  is verified.  

\section{Theory of cosmological perturbations} \label{sec:perturbations}

A major success of inflation is that it provides a mechanism for the generation of density perturbations that seed all the structures in the Universe.  The classical density perturbations originate naturally from quantum fluctuations that grow and become classical due to the exponential expansion.  

The theory of cosmological perturbations describes how the scalar field and the metric fluctuations evolve during inflation.  At the linear level, the homogeneous metric is perturbed by $\delta g_{\mu \nu} $,
\begin{equation} g_{\mu \nu} (\mathbf x) = g_{\mu \nu} ^{\mathrm{FLRW}} +  \delta g_{\mu \nu} (\mathbf x)~.
\end{equation}
There are 10 degrees of freedom (d.o.f.) associated to the metric perturbation $\delta g_{\mu \nu} $.  They can be decomposed in 4 scalar d.o.f. $A,B,C,E$, 4 vector d.o.f. $ B_i $ et $ E_i $ resulting from two space-like vectors of null divergence, 2 tensor d.o.f. $h_{ij}$ resulting from a space-like tensor with vanishing trace and divergence.  
The perturbed metric then can be rewritten as
\begin{equation}
 ds^2 =  a^2 (\eta) \left\{  -( 1+2A ) \dd \eta^2 + 2 (\partial_{i}  B + B_i) \dd x^i \dd \eta 
+ \left[  ( 1 +2 C ) \delta_{ij} + 2  \partial_{i} \partial_{j} E + 2 \partial_{(i} E_{j)} + 2 h_{ij} \right] \dd x^i \dd x^j  \right\} .
 \end{equation}

\subsubsection{The gauge freedom}

One defines a local perturbation of a quantity $Q$  as 
 \begin{equation}
 \delta Q (\mathrm x,t) = Q (\mathrm x, t ) - \bar Q (t )~,
 \end{equation}
 where $\bar Q (t)$ denotes this quantity in the un-perturbed homogeneous space-time.  
 Any perturbation therefore depends on the choice of the coordinate systems on each manifold.   
 In other words, if a coordinate system is fixed for the un-perturbed space-time, one needs to define an isomorphism identifying the points of same coordinates in the two space-times.   The liberty in this choice implies that four d.o.f. are non-physical and only linked to the choice of the coordinate systems on the two manifolds.  

Let us consider a transformation of the coordinate system
\begin{equation}
x^\mu \rightarrow x^\mu + \xi ^\mu~,
\end{equation}
where $\xi^\mu$ is a space-time like vector.   $\xi^\mu$ can be decomposed in two scalar ($T$ and $L$) and two vector ($L_i$) d.o.f.  via
\begin{equation} 
\xi^0 = T~, \hspace{2cm} \xi^i = D^i L + L^i~, \hspace{2cm} D^i L_i = 0~,
\end{equation}
where $D_i $ is defined as the spatial part of the covariant derivative.  Fixing this transformation is thus equivalent in fixing 4 d.o.f..   Under this transformation of the coordinate system, the metric perturbation transforms as 
\begin{equation}
\delta g_{\mu \nu} \rightarrow \delta g_{\mu \nu} + \mathcal L _\xi g_{\mu \nu},
\end{equation}
where $\mathcal L_\xi$ is the Lie derivative along $\xi$.  The Lie derivative evaluates the change of a tensor field along the flow of a given vector field.   It is defined as
\begin{equation} 
\mathcal L_\xi T^{\mu_1 \ldots \  \mu_p}_{\nu_1 \ldots \ \nu_q } = \xi^\sigma \partial_\sigma T^{\mu_1 \ldots \  \mu_p}_{\nu_1 \ldots \ \nu_q }
- \sum_{i=1} ^p T^{\mu_1 \ldots \ \sigma \ldots \ \mu_p}_{\nu_1 \ldots \ \nu_q } \partial_\sigma \xi^{\mu_i} 
+ \sum_{j=1}^q T^{\mu_1 \ldots \ \mu_p}_{\nu_1 \ldots  \ \alpha \ldots \  \nu_q } \partial^\alpha \xi_{\nu_j} .
\end{equation} 
Applied to the symmetric metric, the Lie derivative gives
\begin{equation}
\mathcal L_\xi g_{\mu \nu} = \nabla_\mu \xi_\nu + \nabla_\nu \xi_\mu~,
\end{equation}
where $\nabla_\mu $ is the covariant derivative associated with the metric $g_{\mu \nu} $.  As a result, one can show that scalar, vector and metric perturbations transform like~\cite{PeterUzan}
\begin{eqnarray}
A & \rightarrow & A + T' + \mathcal H T~, \\
B & \rightarrow & B - T + L'~, \\
C & \rightarrow & C + \mathcal H T~, \\
E & \rightarrow & E' + L~, \\
E^i & \rightarrow & E^i + L^i~,\\
B^i & \rightarrow & B^i + {L^i} '~, \\
h_{ij} & \rightarrow & h_{ij}~.
\end{eqnarray}
In the same way, the perturbation $\delta Q$ becomes
\begin{equation}
\delta Q \rightarrow \delta Q + \mathcal L_\xi Q~.
\end{equation}
A quantity is called \textit{gauge invariant} when it is conserved by the transformation of the coordinate system, i.e. when its Lie derivative vanishes. 
Gauge invariants are for instance the Bardeen variables~\cite{PhysRevD.22.1882}
\begin{eqnarray}
\Phi & \equiv & A + \mathcal H (B-E') + (B-E')'~, \\
\Psi & \equiv & -C - \mathcal H (B-E')~.
\end{eqnarray}
If we set $T = B- E'$, $L = -E $ and $L_i' = -B_i$, one has
\begin{equation}
B=E=0~, \hspace{15mm} B_i = 0~,
\end{equation}
and the scalar metric perturbations are identified with the Bardeen variables
\begin{eqnarray}
 A & = & \Phi ,\\
 C & = & -\Psi~.
\end{eqnarray}
This choice is called the \textit{longitudinal gauge}, which we adopt from now.

\subsubsection{Scalar perturbations}

Once the gauge is fixed, one can study the evolution of cosmological perturbations in the linear regime.  Interestingly, scalar and metric perturbations decouple and thus one can consider them separately.   We will show that vector perturbations decay quickly and can be safely neglected.   Let us consider first the scalar perturbations in the longitudinal gauge. The perturbed metric is of the form
\begin{equation}
\dd s^2 = a^2(\eta) \left[ -(1+2 \Phi ) \dd \eta^2 + (1 - 2 \Psi ) \delta_{ij} \dd x^i \dd x^j \right] .
\end{equation}
The scalar field filling the Universe at a given space-time point has an homogeneous part $ \bar \phi $ plus a small perturbation $\delta \phi \ll \bar \phi $,
\begin{equation}
\phi (\mathrm x,t) = \phi  (t) + \delta \phi(\mathrm x,t)~.
\end{equation}
In the longitudinal gauge, it is identified to the gauge invariant variable
\begin{equation}
\delta \phi_{\rr{g.i.}} = \delta \phi + \phi' ( B - E')~.
\end{equation}
After perturbing the energy momentum tensor, the $(0,0)$ and $(i,i)$ first order perturbed Einstein equations read
 \begin{eqnarray} \label{eq:ep1} 
 - 3 \mathcal H (\Psi'+\mathcal H \Phi ) + \nabla ^2 \Psi & = & \frac {4 \pi}{m_{\mathrm p}^2} \left( \phi' \delta \phi' - \phi'^2 \Phi +
a^2 \frac{\dd V}{\dd \phi} \delta \phi \right) ~, \\  \label{eq:ep2}
 \Psi' + \mathcal H \Phi & = & \frac{4\pi}{m_{\mathrm p}^2 } \phi' \delta \phi ~, \\ \label{eq:ep3}
 \Psi'' + 2 \mathcal H \Psi' + \mathcal H \Phi' +& &  
  \Phi \left( 2\mathcal H ' + \mathcal H ^2 \right) + \frac 1 2 \nabla ^2 (\Phi - \Psi ) \nonumber \\
& = & \frac {4 \pi}{m_{\mathrm p}^2}  \left( \phi' \delta \phi' - \phi'^2 \Phi - a^2 \frac {\dd V}{\dd \phi} \delta \phi \right) ~,
\end{eqnarray}
where a prime denotes derivative with respect to the conformal time $\eta$, and where $\mathcal H \equiv a'/a = a H$.  Moreover, because $ \delta T_{i}^{j} \propto \delta_{i}^{j} $ in absence of vector perturbations, one has  $ \Phi = \Psi $.  
On the other hand, the perturbed Klein-Gordon equation reads
\begin{equation}\label{kgp}
\delta \phi '' + 2 \mathcal H \delta \phi' - \nabla^2 \delta \phi + a^2 \delta \phi \frac {d^2V}{\dd\phi^2}
= 2 (\phi'' + 2 \mathcal H \phi' ) \Phi + \phi' ( \Phi' +3 \Psi')~.
\end{equation}
One sees that $ \delta \phi $ is directly related to $\Phi $ and its derivative, so that there remains only one scalar d.o.f..  
By combining Eq.~(\ref{eq:ep1}) and Eq.~(\ref{eq:ep3}), by using Eq.~(\ref{eq:ep2}) as well as the background equations, and after a bit of algebra, an unique second order evolution equation for scalar perturbations can be derived,
\begin{equation} \label{eq:Phievol}
\Phi'' + 2 \left( \mathcal H - \frac {\phi''}{\phi'} \right) \Phi' - \nabla^2 \Phi + 2 \left( \mathcal H' - \mathcal H \frac {\phi''}{\phi'} \right) \Phi = 0~.
\end{equation}
It is convenient to work in Fourier space, because in the linear regime each mode evolves independently and one can solve the evolution equation for each of them.  After a Fourier expansion, one can define
\begin{eqnarray}
\mu_{\mathrm s} & \equiv &  -  \frac {4 \sqrt \pi}{  m_{\mathrm p}}  a (\delta \phi + \phi' \Phi / \mathcal H )~,\\
 \omega_{\mathrm s} ^2 & \equiv & k^2 - \frac {(a\sqrt\epsilon) '' }{a\sqrt \epsilon}~,
\end{eqnarray}
where $k$ is a comoving Fourier wavenumber, and equation (\ref{eq:Phievol}) can be rewritten in a simpler form, 
\begin{equation} \label{pertscal}
 \mu_{\mathrm s}'' + \omega_{\mathrm s} ^2 (k,\eta) \mu_{\mathrm s} = 0~.
\end {equation}
This equation is similar to an harmonic oscillator with a varying frequency.  Apart in some specific cases, it cannot be solved analytically and one has to use numerical techniques.  It is also possible to solve it analytically after a Taylor expansion at first order in slow-roll parameters.  

Instead of $\Phi$ or $\mu_{\rr s} $, it is a common usage to calculate the mode evolution and the power spectrum of the curvature perturbation $\zeta $\footnote{$ \zeta$ can be identified to the spatial part of the perturbed Ricci scalar in the comoving gauge, in which the fluids have a vanishing velocity ($\delta T^0_{\ i} = 0 $).} defined as
\begin{equation} \label{zeta}
\zeta \equiv  \Phi - \frac{\mathcal H}{\mathcal H' - \mathcal H^2 } (\Phi' + \mathcal H \Phi )= - \mu_{\mathrm s} \frac 1 {2a\sqrt {\epsilon_{\rr 1}} }~.
\end{equation} 
Its power spectrum thus reads
\begin{equation}
 \mathcal P_{\zeta} (k) = \frac {k^3}{8\pi ^2 } \left| \frac {\mu_{\mathrm s}}{a\sqrt { \epsilon_1} } \right| ^2~.
 \end{equation}
By using Eq.~(\ref{kgp}), one can determine that $\zeta$ evolves according to
\begin{equation}
\zeta' = \frac{- 2 \mathcal H}{3 (1+w)}  \left( \frac{k}{\mathcal H}  \right)^2 \Phi~,
\end{equation}
and as long as the modes are super-Hubble ($ k/\mathcal H \ll 1 $), $\zeta (k) $  remains constant in time.   Therefore, observable modes re-entering into the Hubble radius during the matter dominated era have kept the value they had during inflation, when they exit the Hubble radius, independently of the details of the reheating phase\footnote{Let notice that a non linear growth of density perturbations during preheating is expected in some models, possibly affecting the linear curvature perturbations on very large scales~\cite{Jedamzik:2010dq,Jedamzik:2010hq}.} and the transition between inflation and the radiation dominated era.  For 1-field inflationary models, they can be used to probe directly the inflationary era.  

 \begin{figure}[htbp] 
	\begin{center}
	\includegraphics[height=80mm]{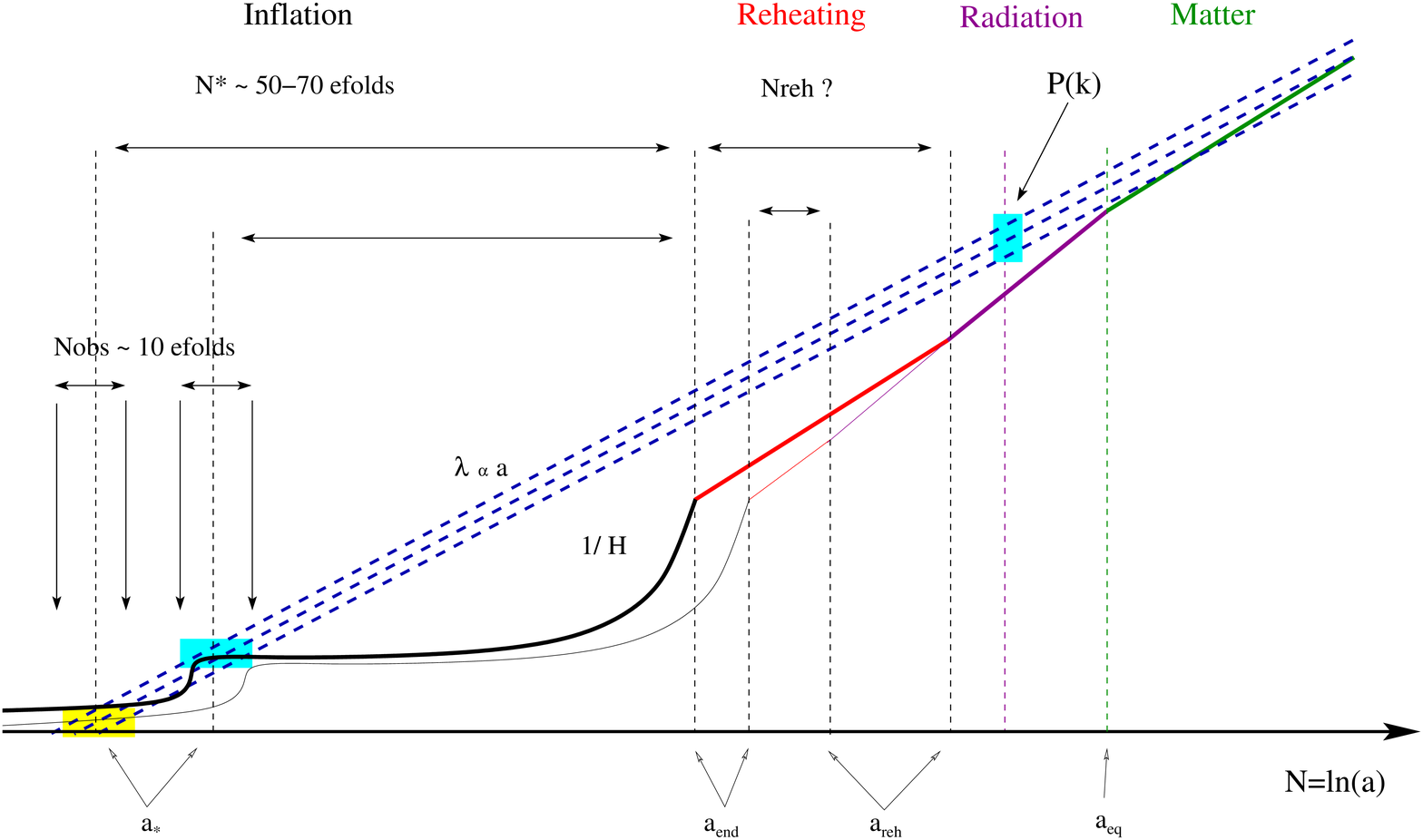}
	
	\caption{This scheme form Ref.~\cite{Ringeval:2007am} illustrates how observable perturbation modes evolve during and after inflation.  The horizontal axis represents the number of e-folds generated from the onset of inflation.  Observable modes exit the Hubble radius on a range of about ten e-folds.  From this time, inflation still lasts from 50 to 70 e-folds, depending on the energy scale of inflation~\cite{Liddle:2003as} and on the duration of the reheating phase.  The curvature and tensor perturbations are constant for super-Hubble wavelengths, until they re-enter into the Hubble radius during the matter/radiation dominated era. }
	\label{modes}
	\end{center}
\end{figure}

\subsubsection{Tensor perturbations}

The metric for the tensor perturbations reads
\begin{equation}
 \dd s^2 = a^2(\eta) \left[ - \dd \eta^2 + (1 + h_{ij})  \dd x^i \dd x^j \right]~,
\end{equation}
and the metric perturbation $h_{ij}$ is gauge invariant.  It is convenient to express the two d.o.f. in $h_{ij}$ as
\begin{equation}
 h_{ij} = a^2 \left(
\begin{array}{lll}
h_+ & h_\times & 0 \\
h_\times & h_+  & 0 \\
0 & 0 & 0\\  \end{array} \right).
\end{equation}
As for the scalar perturbations, one can then write the first order perturbed Einstein equations,
\begin{equation}
h_{\alpha}'' + 2 \mathcal H h_{\alpha} + \nabla^2 h_{\alpha} =0,
\end{equation}
where $ \alpha = +,\times $.  By defining 
\begin{equation}
 \mu_{\mathrm t} \equiv \frac 1 2 ah_{ij} \delta^{ij} ,
 \end{equation}
after Fourier expansion, these two equations reduce to
\begin{equation} \label{tensoriel}
 \mu_{\mathrm t} '' + \omega_{\mathrm t} ^2  (k,\eta) \mu_{\mathrm t} = 0 ~,
 \end{equation}
where
 \begin{equation}
  \omega_{\mathrm t}^2 (k,\eta) \equiv k^2 - \frac {a'' }{a} ~.
  \end{equation}
The variable $ h = h_{ij} \delta^{ij}, $ is the analogous of $\zeta$ for the tensor perturbations and has similar properties.  Its power spectrum reads
\begin{equation}
\mathcal P_{h} (k) = \frac { 2 k^3 }{ \pi^2} \left| \frac {\mu_{\mathrm t} }{a} \right| ^2 .
 \end{equation}
 
 \subsubsection{Vector perturbations}
 
 The metric for the vector perturbations in the longitudinal gauge reads
 \begin{equation}
 \dd s^2 = a^2 (\eta) \left[  - \dd \eta^2 + 2 \partial_{(i} E_{j)} \dd x^i \dd x^j  \right]~,
 \end{equation}
and the vector perturbations $E_j $ can be identified in this gauge to the gauge invariant variable
\begin{equation}
 \Phi_i = E_i - B_i~.
\end{equation}
The perturbed energy-momentum tensor for a scalar field does not contain any source of vector perturbations and the first-order perturbed Einstein equations read
\begin{equation}
\Phi_i'' + 2 \mathcal H \Phi_i' = 0~.
\end{equation}
Vector perturbations therefore decay quickly, since $\Phi_i' \propto a^{-2} $ and because $a$ grows nearly exponentially with the cosmic time.  That is why vector perturbations are  usually neglected.

\subsubsection{Quantification of perturbations}

In the context of inflation, quantum fluctuations are responsible for large scale structures of the Universe observed today.   The canonical commutation relations are the basis of the quantization process.  But to define them, one needs the canonical momenta, and thus the action.  It is incorrect to interpret directly the classical equations of motion [Eqs.~(\ref{pertscal}) and~(\ref{tensoriel})] quantum mechanically, because it leads in general to an incorrect normalization of the modes~\cite{Mukhanov:1990me}. 


\paragraph{Scalar perturbations:}
If we perturb the total action of the system up to the second order in the metric and scalar field perturbations, one finds~\cite{Mukhanov:1990me}
\begin{equation}
^{(2)} \delta S = \frac 1 {2} \int \dd^4 x \left[ (v')^2 - \delta^{ij} \partial_i v \partial_j v + \frac{z_{\rr s}''}{z_{\rr s}} v^2 \right]~.
\end{equation}
where $ v \equiv a ( \delta \phi_{\rr{g.i.}} + \phi' \Phi) / \mathcal H $ can be identified to $- \mu_{\rr s} \mpl / 4 \sqrt \pi $ in the longitudinal gauge, and is the so-called \textit{Mukhanov-Sasaki variable}.    The quantity $z_{\rr s}$ is defined as $z_{\rr s} \equiv \sqrt{4 \pi} a \phi' / \mathcal H = a \sqrt \epsilon_1 $.  As expected, the e.o.m for this Lagrangian reads
\begin{equation} \label{eq:class_evolu}
v'' - \left( \nabla^2 + \frac{z''}{z}  \right) v = 0~.
\end{equation}
The first step of the quantization process is to determine $\pi $, the conjugate of $v$, 
\begin{equation} 
\pi = \frac{\delta \mathcal L}{\delta v'} = v'~.
 \end{equation}
Then the Hamiltonian reads 
\begin{equation}
H = \int \dd x ^4 \left( \pi^2 + \delta^{ij} \partial_i v \partial_j v - \frac{z_s''}{z_s}  v^2  \right) ~.
\end{equation}
In a quantum description, the classical variables $v$ and $\pi$ are promoted as quantum operators $\hat v$ and $ \hat \pi $, satisfying the commutation relations 
\begin{equation}
[\hat v (\mathbf x, \eta ) ,\hat v (\mathbf y, \eta ) ] = [\hat \pi (\mathbf x, \eta ) ,\hat \pi (\mathbf y, \eta ) ] = 0~,
\end{equation}
\begin{equation}
[\hat v (\mathbf x, \eta ) ,\hat \pi (\mathbf y, \eta ) ] = i \delta^{(3)} (\mathbf x - \mathbf y) ~.
\end{equation}
In the Heisenberg picture, the operator  $\hat v$ can be expanded over a complete orthonormal basis of the solution of the field equation Eq.~(\ref{eq:class_evolu}).  If one takes a basis of plane waves, one has
\begin{equation}
\hat v (\mathbf x, \eta) = \frac{1}{(2 \pi)^{3/2}}\int \dd^3 \mathbf k \left( v_k \rr e^{i  \mathbf {k\cdot x} } \hat a_{ \mathbf k} + v_k^* \rr e^{-i  \mathbf {k\cdot x} } \hat a_{ \mathbf k}^+   \right)~,
\end{equation}
and the equation for the $v_k(\eta)$ is 
\begin{equation} \label{eq:vk}
v_k'' (\eta) + \left(k^2 -   \frac{z''}{z} \right) v_k =0~. 
\end{equation}
If the normalization condition 
\begin{equation}
v_k' (\eta) v_k^* (\eta) - {v_k^*}' (\eta) v_k (\eta) = 2 i
\end{equation}
is satisfied, the creation and annihilation operators $  \hat a_{ \mathbf k} $ and $ \hat a_{ \mathbf k}^+ $ satisfy the standard commutation relations 
\begin{equation}
[\hat a _{\mathbf k },\hat a _{\mathbf k' } ]=[\hat a _{\mathbf k }^+ ,\hat a _{\mathbf k' }^+ ] = 0~, \hspace{15mm} [\hat a _{\mathbf k },\hat a _{\mathbf k' }^+ ] = \delta^{(3)} (\mathbf {k - k'} )~.
\end{equation}
At a time $\eta_{\rr i} $, the vacuum $| 0 \rangle $ can now be defined, such that for all $\mathbf k$ one has
\begin{equation}
 \hat a_{\mathbf k} | 0 \rangle = 0~.
\end{equation}
From Eq.~(\ref{eq:vk}), in the sub-Hubble regime, we have
\begin{equation} \label{condinit_scalar}
\lim_{k/aH \rightarrow + \infty} v_k (\eta) = \frac { \mathrm e^{-ik(\eta-\eta_{\mathrm i}) } }{\sqrt{2k} } .
\end{equation}
This can be used to give consistent initial conditions to Eq.~(\ref{pertscal}).  

\paragraph{Tensor perturbations:}
 The quantification of the tensor perturbations is analogous.  One can first determine the second order perturbed action
 \begin{equation}
 ^{(2)} \delta S = -  \frac{\Mpl^2}{2} \sum_{\alpha=+,\times} \int \dd^4 x \left[ (h_\alpha')^2 - \delta^{ij} \partial_i h_\alpha \partial_j h_\alpha + \frac{a''}{a} h_\alpha^2 \right]~.
 \end{equation}
 The perturbations $h_\alpha (\eta, \mathbf x ) $ are the canonical variables.  They are promoted as quantum operators and are expanded in plane waves, 
 \begin{equation}
\hat h_j (\mathbf x, \eta) = \frac{1}{(2 \pi)^{3/2}}\int \dd^3 \mathbf k \left( h_{k,j} \rr e^{i  \mathbf {k\cdot x} } \hat a_{ \mathbf k , j} + h_{k,j}^* \rr e^{-i  \mathbf {k\cdot x} } \hat a_{ \mathbf k,j}^+   \right)~.
\end{equation}
The e.o.m. are
\begin{equation}
h''_{\mathbf k,j} + \left(  k^2 - \frac{a''}{a} \right) h_  {\mathbf k,j} = 0~,
\end{equation}
similar to Eq.~(\ref{tensoriel}).  
The quantification process can be used to determine the sub-Hubble tensor perturbation evolution,
\begin{equation} \label{condinit_tensor}
\lim_{k/aH \rightarrow + \infty} h_{k,j} (\eta) =\frac { \mathrm e^{-ik(\eta-\eta_{\mathrm i}) } }{\sqrt{2k} }~.
\end{equation}

\subsubsection{Expansion in slow-roll parameters} \label{sec:sr_expand}

Eqs.~(\ref{pertscal}) and~(\ref{tensoriel}) can be solved analytically if these are expanded at first order in the Hubble flow-functions around some pivot scale.
To do so, let us first rewrite 
\begin{equation} \label{eq:etasr}
\eta = \int \frac {\dd t}{a } = \int \frac{\dd a}{H a^2} = - \frac 1 {aH} + \int \dd a \frac {\epsilon_1}{a^2 H}~.
\end{equation}
In the slow-roll approximation, one has $|\epsilon_i| \ll 1$.  By definition, the derivative of the first and second Hubble-flow functions with respect to the number of e-folds are second order in $|\epsilon_i|$,
\begin{equation}
\frac{\dd \epsilon_1}{\dd N} = \epsilon_1 \epsilon_2~,\hspace{10mm} \frac{\dd \epsilon_2}{\dd N} = \epsilon_2 \epsilon_3~.
\end{equation}   
One can therefore neglect their variation over the time taken for observable modes to exit the Hubble radius (it corresponds typically to $\Delta N \sim 10$~\cite{Liddle:2003as}).   In this approximation, and by using Eq.~(\ref{eq:etasr}), one thus has
\begin{eqnarray}
a H & = & - \frac 1 \eta + \frac{a H} {\eta} \int \frac{\epsilon_1 } {a^2 H} \dd a  =  - \frac 1 \eta + \frac{a H} {\eta} \int \frac{\epsilon_1 } {a} \dd t \simeq  - \frac 1 \eta + a H \epsilon_1 \simeq  \frac{-1 + \epsilon_1} {\eta}~.  \label{eq:aH}
 \end{eqnarray}
By integrating the last expression, the scale factor is found to behave like
\begin{equation} \label{eq:aeta}
a(\eta) \simeq l_0 | \eta | ^{-(1+\epsilon_1)} \simeq  \frac{l_0}{|\eta|} \left( 1 - \epsilon_{1} \ln |\eta| \right) ~,
\end{equation}
where $l_0$ is an arbitrary parameter.  
Instead of choosing an arbitrary scale, it is more convenient to chose an arbitrary conformal time $\eta_* $, and to relate it to the scale $l_0 $ via
\begin{equation} \label{eq:Hetastar}
H(\eta_*) = - \frac{1 + \epsilon_{1*}}{a \eta_*} \simeq \frac 1 {l_0} \left[ 1 + \epsilon_{1*} (1+ \ln | \eta_* | ) \right] \equiv H_*~,
\end{equation}
where a star subscript denotes the evaluation at the time $\eta_*$.  We have fixed  $\eta_* $ such that the following relation is verified
\begin{equation}
k_* = a(\eta_*) H(\eta_*)~,
\end{equation}
where $k_*$ is the comoving pivot mode introduced in Section~\ref{sec:powerspectrum}.  

\vspace{2mm}
The scalar and tensor perturbations evolve according to Eq.~(\ref{pertscal}) and Eq.~(\ref{tensoriel}).  These equations can now be expanded at first order in slow-roll parameters and then solved analytically.  The first step is to use Eq.~(\ref{eq:aH}) to expand
\begin{eqnarray}
 \frac { (a \sqrt \epsilon)''}{a \sqrt \epsilon } & \simeq & 
 \frac { 2 + 3 \epsilon_1 + \frac  3 2 \epsilon_2 }{ \eta ^2 } ~, \\
\frac { a'' }{a} & \simeq & \frac { 2 + 3 \epsilon_1 } {\eta^2} ~.
\end{eqnarray}
In the approximation that the slow-roll parameters are constant in time, a general solution to  (\ref{pertscal}) and (\ref{tensoriel}) can be found 
\begin{equation}
 \mu_{\mathrm s,\mathrm t} (k\eta ) = \sqrt {k \eta } \left[ A J_{\nu_{\mathrm s,\mathrm t}} (k\eta ) + B J_{-\nu_{\mathrm s,\mathrm t}} (k\eta) \right] ~,
 \end{equation}
where $ \nu_{\mathrm s} = - \frac  3 2 - \epsilon_1 - \frac 1 2 \epsilon_2  $ and $ \nu_{\mathrm t} = - \frac 3 2 - \epsilon_1 $.
It is convenient to express the Bessel function $J_{\nu} (k\eta ) $ in terms of the Hanckel functions of first and second kind $H^{(1)} _\nu (k\eta)$ et $ H^{(2)}_\nu(k\eta)$.
The quantification of the perturbations provide the initial conditions.  By using the asymptotic behavior of the Hanckel functions,
\begin{eqnarray}
H_\nu ^{(1)} (z \rightarrow \infty ) & = & \sqrt{ \frac 2 {\pi z} } \rr e^{i(z-\frac 1 2 \nu \pi - \frac 1 4 \pi)} , \\
H_\nu ^{(2)} (z \rightarrow \infty ) & = & \sqrt{ \frac 2 {\pi z} } \rr e^{-i(z-\frac 1 2 \nu \pi - \frac 1 4 \pi)} ,\\
\end{eqnarray}
and by comparing with the Eqs.~(\ref{condinit_scalar}) and (\ref{condinit_tensor}), $A$ and $B$ can be determined.  For scalar perturbations, one has
\begin{eqnarray} A & = & 2 i \frac \pi {\mpl} \sqrt k \sin (\pi \nu_s) \rr e^{i\left( \frac 1 2 \nu_s - \frac \pi 4 + k \eta_{\rr i} \right)}~,\\
B & = & - A \rr e^{-i \pi \nu_s}~.
\end{eqnarray}
On the other hand, one can use the limit condition
\begin{equation}
H^{(1)} _{1/2-\nu} (z \rightarrow 0) = - \frac i \pi \Gamma \left( \frac 1 2 - \nu \right) \left( -\frac z 2 \right)^{\nu - \frac 1 2} ,
\end{equation}
as well as the recurrence relation
\begin{eqnarray}
\Gamma (z+\epsilon) & = & \epsilon \psi(z) \Gamma(z) + \Gamma(z)~,\\
\psi(1/2) & = & - \gamma_{\mathrm{Euler}} - 2 \ln 2~, 
\end{eqnarray}
with $  \gamma_{\mathrm{Euler}} \simeq 0.5772 $ and where $\psi(z)$ is the polygamma function, to obtain the super-Hubble behavior of the perturbation modes. 
Since observable modes are super-Hubble at the end of inflation, one obtain the power spectrum expanded at first order in slow-roll parameters around $\eta_*$, by using Eqs.~(\ref{eq:aeta}) and~(\ref{eq:Hetastar}).  For scalar perturbations, one obtains   
\begin{eqnarray}
\mathcal P_{\zeta}(k) & = &  \frac {k^3} {8\pi ^2 } \left| \frac {\mu_{\mathrm s}}{a\sqrt { \epsilon_{1*}} } \right| ^2 \\
& = & \frac {H_* ^2} {\pi m_{\mathrm p} ^2 \epsilon_{\rr 1 *} } \left[ 1 - 2 (C+2) \epsilon_{1*} + C \epsilon_{2*}  - (2 \epsilon_{1*} + \epsilon_{2*} )  \ln \left( \frac {k}{k_*} \right)  \right]~.
\end{eqnarray}
For tensor perturbations,
\begin{eqnarray}
 \mathcal P_{h}(k) & = & \frac { 2  k^3 }{\pi^2} \left| \frac {\mu_{\mathrm t} }{a} \right| ^2 \\
 & = & \frac { 16 H_* ^2 }{\pi m_{\mathrm p}^2 } \left[ 1- 2(C+1) \epsilon_{1*} - 2 \epsilon_{1*} \ln \left( \frac k {k_*} \right) \right]~,
\end{eqnarray}
with $C = \gamma_{\mathrm{Euler}} + 2 \ln 2 - 2 $.  All the slow-roll parameters are evaluated at $\eta_*$.
At first order in slow-roll parameters, the scalar spectral index is therefore
\begin{equation}
n_{\mathrm s} -1 = - 2 \epsilon_{\rr 1*} - \epsilon_{\rr 2*}~.
\end{equation}
For the tensor perturbations, it is 
\begin{equation}
n_{\mathrm t}  = - 2 \epsilon_{1*}~.
\end{equation}
Finally the ratio between the tensor and scalar power spectrum is given by
\begin{equation}
r = 16 \epsilon_{\rr 1 *}~.
\end{equation}
The amplitude and the spectral tilt of the scalar and tensor power spectra can thus be derived easily in the slow-roll approximation, at first order in slow-roll parameters, for a given scalar field potential.   They are given only in terms of the potential and its derivatives with respect to the scalar field.

Finally, note that $r = - 8 n_{\mathrm t} $ is a generic prediction of single field inflation.  Checking this relation will be a major goal for future experiments.  If observations show that it is satisfied, this should be seen as a proof that inflation really took place, since most of the alternatives predict different behaviors.  This relation is called  the \textit{consistency relation of inflation}.

\section{Worked example: the power-law large field potential} \label{sec:largefield}

To illustrate the results of the two previous sections, let consider one of the simplest potential,  of the power-law form (the model is often referred as large field inflation or chaotic inflation):
\begin{equation} \label{eq:largefield}
V(\phi) = M^4 \left( \frac { \phi}{M_{\mathrm p}}  \right) ^{p}~.
\end{equation}
The background dynamics in the slow-roll approximation is given by Eqs.~(\ref{sr1}) and (\ref{sr2}).   The number of e-folds realized from an initial field value $ \phi_{\mathrm i} $ can be determined analytically, 
\begin{equation} N(\phi) = \frac {1 }{ 2 p} \left[ \left( \frac {\phi_{\mathrm i}}{M_{\mathrm p}} \right)^2 - \left(\frac {\phi}{M_{\mathrm p}} \right) ^2 \right]~. \end{equation}
The first and second slow-roll parameters read
\begin{equation}
\epsilon_1 (\phi) = \frac {p^2 M_{\mathrm p}^2 }{2 \phi^2 } ~,
\end{equation}
\begin{equation}
 \epsilon_2 (\phi)= \frac { 2 p M_{\mathrm p} ^2 }{  \phi^2 } ~.
 \end{equation}
Inflation stops when the first slow-roll parameter reaches $ \epsilon_1 = 1$.  This corresponds to the inflaton value
\begin{equation} 
\frac {\phi_{\mathrm {end}}}{M_{\mathrm p}} = \frac p {\sqrt 2} ~. 
\end{equation} 
For a given number of e-folds $N_*$ between the Hubble exit of the pivot mode and the end of inflation, the inflaton value $\phi_*$ and the slow-roll parameters can be obtained in a straightforward way.  For $N_* = 60$ and $p=2$, they read
\begin{eqnarray}
\phi_{*} & = &  \sqrt{ 2 p \left(N_* + \frac p 4  \right)} M_{\rr{p}} \simeq 15.5 \Mpl \simeq 3.1 m_{\rr p} ~,\\
 \epsilon_{1*} & \simeq & 0.0083~, \hspace{10mm} \epsilon_{2*} \simeq 0.0166~.
\end{eqnarray}
It is then straightforward to derive the scalar power spectrum spectral index and the scalar to tensor ratio,
\begin{equation}
 n_{\rr s} =  1- 2 \epsilon_{1*} - \epsilon_{2*} \simeq  0.967~,     \hspace{10mm} r = 16 \epsilon_{1 *} \simeq 0.13~.
\end{equation}
These predictions are independent of the mass of the field and correspond to a point in the $(n_{\rr s}, r)$ plane.   Nevertheless, they depend on the reheating history through $N_*$.  The mass scale is fixed by the scalar power spectrum amplitude given in section~\ref{sec:observables}.  One gets $M \simeq 10^{-3} \mpl $.    For large field models, inflation takes therefore place close to the GUT scale.  
Let remark that the inflaton field must be initially super-Planckian in order for inflation to last at least 60 e-folds.  Nevertheless, the energy density remains much smaller than the Planck scale. General Relativity is thus valid and no effect of quantum gravity is expected.  

 \begin{figure}[htbp] 
	\begin{center}
	\includegraphics[width=150mm]{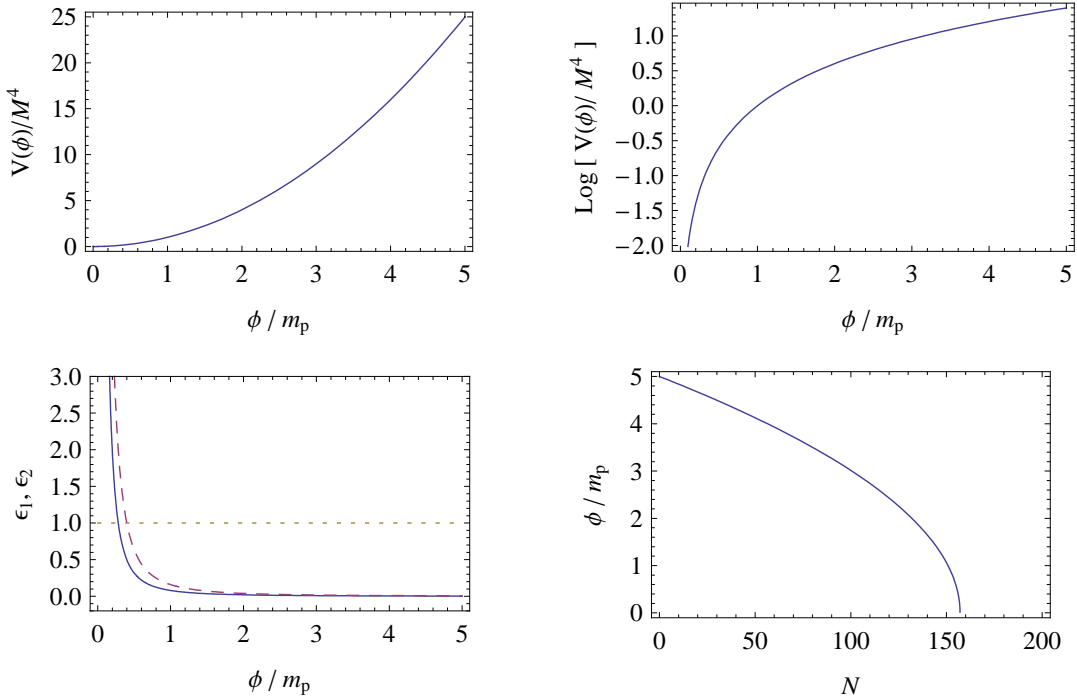}
	
	\caption{Top:  Scalar field potential and its logarithm for the large field model and for $p=2$.  Bottom-left:  evolution of slow-roll parameters  $\epsilon_1$ (solid line) and $\epsilon_2$ (dashed line).  Inflation stops when $\epsilon_1 = 1$ (dotted line).  Bottom-right: evolution of $\phi(N)$, for an initial field value $ \phi_{\mathrm i}= 5 \mpl $}
	\label{largefield}
	\end{center}
\end{figure}

\section{Classification of single-field potentials} \label{sec:models}

There exists a large variety of single-field models of inflation, from the simplest power-law potentials to more complicated potentials arising in various high energy frameworks.  For an exhaustive list of models and their analysis, one can refer to the \textit{Encyclopaedia Inflationaris}~\cite{Martin:2013tda}.  Note that all those models are implemented numerically within the ASPIC library.  

In this section, I introduce a classification of single-field potentials, depending on the values taken by the scalar field and the slow-roll parameters.  It is closed to the Schwartz-Terrero-Escalante classification~\cite{Schwarz:2004tz,Martin:2013tda}.  Note however that it is not really complete since more complicated forms of the potential could be envisaged.  Moreover, some scalar field potentials can belong to several classes depending on their parameters.  

\begin{enumerate}
\item \textbf{Small-field models:}  Slow-roll parameters are usually such that $0 < \epsilon_1 \ll \epsilon_2 $, meaning that the kinetic energy increases, as well as the ratio between the kinetic energy and the total energy.  The Higgs model as well as plateau potentials, e.g. of the form $V\propto [1- (\phi/ \Mpl)^p ]$, belong to this class.  The low value of $\epsilon_1$ ensures that many e-folds can be realized close to $\phi = 0$ at sub-Planckian field values.  The spectral index deviates from unity due to the concavity of the potential, which is negative (since $\epsilon_2 \sim - V''/V$).
\item  \textbf{Large-field models:}  In this class of models, one has usually $2 \epsilon_1 \gtrsim \epsilon_2 > 0$, meaning that the ratio between kinetic energy and the total energy increases whereas the kinetic energy decreases.  The power-law potential ($V \propto \phi^p$) belongs to this class.  The spectral index deviates from unity mostly due to the $\epsilon_1$ parameter, and given the present constraints this implies super-Planckian field values.  Note however that the energy density during inflation is still much lower than the Planck scale.  
\item \textbf{Hybrid models:}  In this class of models, inflation takes place in a false vacuum dominated regime, often along a nearly flat valley of some multi-field potential.   Inflation can end due to the presence of auxiliary fields inducing a tachyonic waterfall instability below some critical field value $\phi_c$, or due to a change in the shape of the potential.   Both the kinetic energy and the ratio between kinetic and total energy decrease.  The potential is e.g. of the form $V\propto [ 1+(\phi / \mu )^p ]$ (original hybrid model) and many e-folds of inflation can be realized in the false vacuum, at sub-Planckian field values.  However in this model one gets $\epsilon_2 < 0$ and $\epsilon_1 \ll 1$, and thus the spectral index is blue, which is now ruled out.  There are nevertheless hybrid models, e.g. SUSY models (F-/D-term models and variants), where a flat direction of the potential is lifted up by logarithmic radiative corrections. This leads to $\epsilon_2 > 0$ and thus a red spectrum possibly in agreement with observations.  Finally note that in hybrid models, multi-field effects can be important, e.g. before field trajectories reach the inflationary valley~\cite{Clesse:2009ur,Clesse:2009zd,Clesse:2008pf}, or during the final waterfall phase~\cite{Clesse:2013jra,Clesse:2012dw,Clesse:2010iz}.
\end{enumerate}

\section{Observational constraints from Planck and BICEP2} \label{sec:observations}

\subsection{Planck results}

\begin{figure}[ht] 
 	\begin{center}
 \includegraphics[height=70mm]{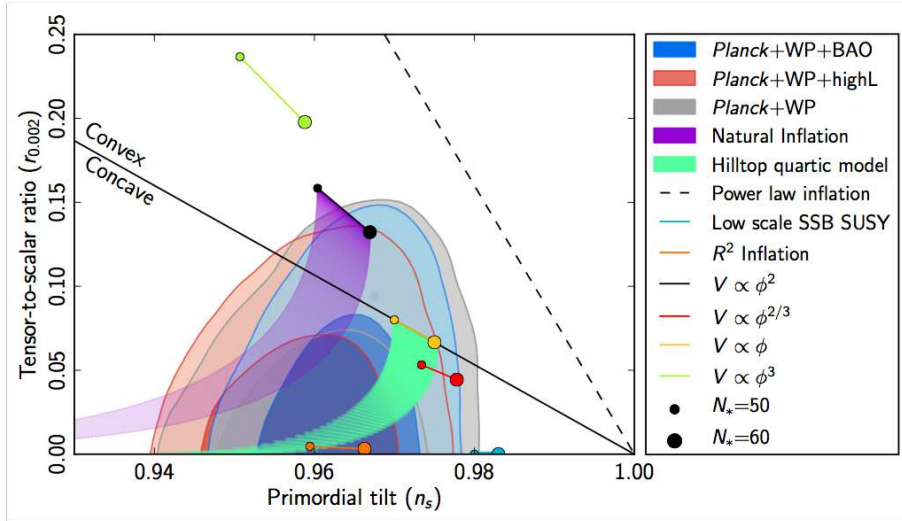}
	\caption{ 1-$\sigma$ and 2-$\sigma$ contours in the plane $(n_{\rr s},r)$ from Planck plus BAO or WMAP polarization~\cite{Ade:2013lta,Planck:2013kta}.  The predictions for several inflation models are also reported. }  
	 \label{fig:Plancknsr}
	\end{center} 
\end{figure}

In 2013 the Planck mission has delivered its measurements of the CMB temperature anisotropies.   The CMB angular power spectrum has been measured with a high accuracy down to very small scales.  It is found to be in a very good agreement with the best fit of the standard cosmological model, assuming that primordial perturbations are well described by the amplitude and the spectral index of their power spectrum.  With Planck combined to observations of the Baryon Acoustic Oscillations and type 1-a supernovae, the cosmological parameters have been measured with an unprecedented accuracy~\cite{Ade:2013lta,Planck:2013kta}.   As already mentioned, the scalar power spectrum amplitude and spectral index are respectively given by $\As = 2.196^{+0.051}_{-0.06} \times 10^{-9} $ and $\ns  = 0.9603 \pm 0.0073$.   An important point is that $n_s <1$ at more than $4\sigma$, which rules out models predicting a scale invariant or a blue tilted scalar power spectrum (such as the original hybrid model).   An upper bound on the tensor to scalar ratio ($r \lesssim 0.11$) has also been derived by combining Planck data to the measurements by WMAP of the CMB polarization.   The Planck limits in the plane $(\ns,r)$ are given in Fig.~\ref{fig:Plancknsr}, together with the predictions of some of the most well-known inflation models.   One can observe that the limit where convex potentials will be disfavored at the 95\% C.L. is not far, and that simple potentials like $V\propto \phi^4$ and $V \propto \phi^3$ are already strongly disfavored.   Simple supersymmetric models like F-term and D-term inflation are also in strong tension with Planck data, because they predict typically that $0.98 \lesssim \ns \lesssim 1$.  

Planck also give the best constraints on the level of local non-Gaussianities, with a bound $\fNL^{\rr{loc}} = 2.7 \pm 5.8  $ \cite{Ade:2013ydc}.  This excludes many multi-field inflation scenarios and tends to favor single-field models.  Planck has also improved the limits on a possible contribution of a cosmic string network to the CMB temperature anisotropies.

We review in more details the implications of the Planck results for inflation models in the next section.  

 \begin{figure}[ht] 
 	\begin{center}
	\includegraphics[height=80mm]{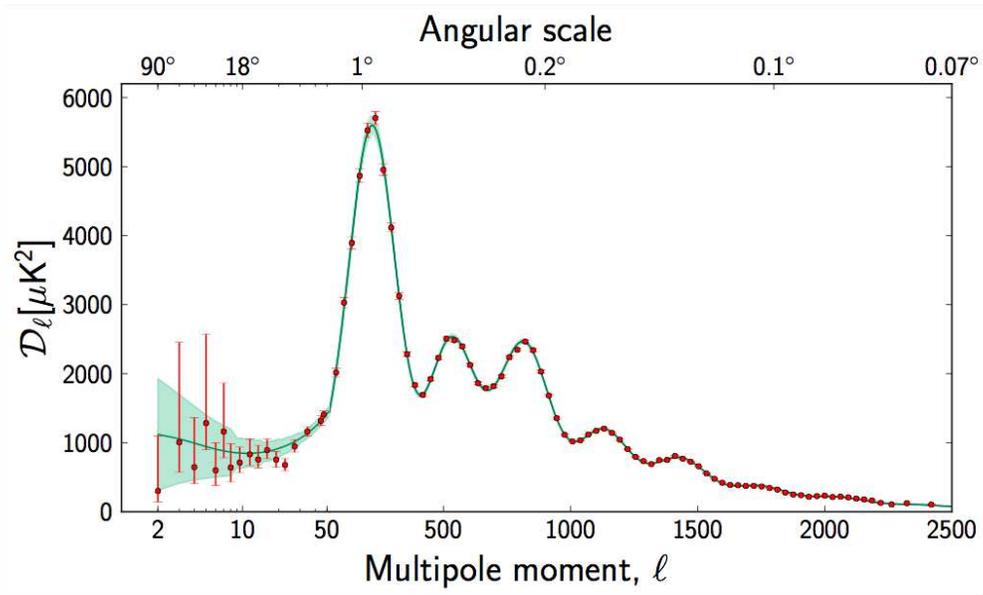}
	\caption{ CMB temperature anisotropy angular power spectrum seen by Planck~\cite{Ade:2013lta,Planck:2013kta}, with the predictions for the best fit of the standard cosmological model parameters, assuming a scalar power spectrum described by an amplitude $A_{\rr s}$ and a spectral index $\ns$.    }  
	 \label{fig:PlanckCls}
\end{center} 
\end{figure}

\subsection{BICEP2 results}

\begin{figure}[ht] 
 	\begin{center}
	\includegraphics[height=70mm]{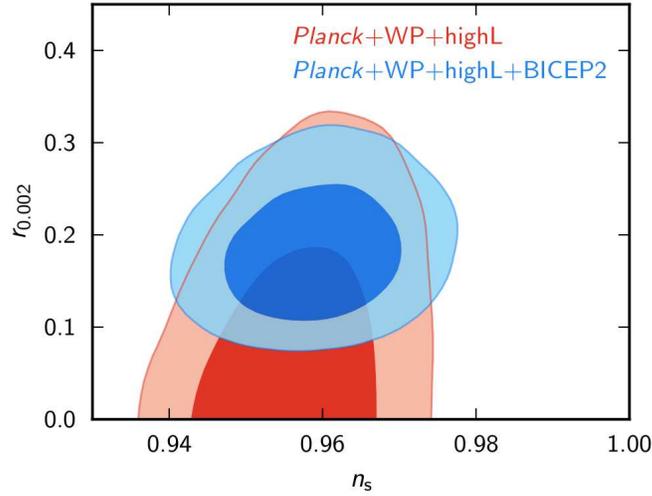}
	\caption{ Planck and BICEP2  combined constraints in the plane $(n_s,r)$.  Figure from~\cite{Ade:2014xna}. }  
	 \label{fig:Bmodespectrum}
\end{center} 
\end{figure}

The BICEP2 experiment is based at the South Pole in Antarctica and measures the CMB polarization on large scales, in a small patch of the sky.   In March 2014 it has reported the measurement of B-mode polarization of the Cosmic Microwave Background~\cite{Ade:2014xna} and claimed that it is of primordial origin.  The signal can be attributed to the tensor modes generated during inflation for a tensor to scalar ratio $r = 0.20^{+ 0.07}_{-0.05} $.  They also claimed to exclude $r = 0$ at more than $7 \sigma$.  Combined Planck and BICEP2 constraints in the plane $(n_s,r)$ are reported on Fig.~\ref{fig:Bmodespectrum}.  Both experiments are in agreement but one can note some tension between them.  As explained in the next section, this is translated in many models of inflation that are incompatible with both Planck and BICEP2. 

The theory of the CMB polarization is rather complex and goes beyond the scope of these lecture notes.  Here I only briefly comment the principal result.   The polarization of the CMB is due to the Compton diffusion that tends to polarize the radiation in the orthogonal direction to the diffusion plane.   In the case of perfect isotropy of the radiation, or in the case of dipolar anisotropy, there is no net effect and the CMB would not be polarized.  However, for quadripolar anisotropies, the CMB radiation becomes polarized in averaged.   On can distinguish two types of polarization, designated by E-mode and B-mode by analogy to the electromagnetism because they correspond to curl free gradient and divergence free curl polarizations.   The important point is that tensor perturbations generate both E-mode and B-mode polarization whereas scalar perturbations generate only E-modes.  By measuring primordial B-modes, one therefore has access to the amplitude of the power spectrum of tensor perturbations, which has been calculated in a previous section.

Measuring the tensor to scalar ratio has profound implications for inflation.  Indeed, the tensor to scalar ratio is directly related to the first slow-roll parameter, $ r = 16 \epsilon_1$.  Knowing the value of $\epsilon_{1*}$ together with the amplitude of the power spectrum of scalar perturbations then permits to fix $H_*$ as well as the value of the field and of the potential at the time of Hubble exit of the pivot scale.  Measuring $r$ therefore gives direct access to the energy scale of inflation.  Combined with measurements of the scalar spectral index, this gives also the curvature of the potential via the second slow-roll parameter $\epsilon_2$.  The energy scale of inflation is given by
\be
\rho^{1/4} \simeq 2.2 \left(   \frac{r}{0.2}\right) 10^{16} \ \mathrm{GeV}
\ee
and therefore BICEP2 results point toward an energy scale of inflation close to the GUT scale.   One can also note that BICEP2 results lead to a relatively large value of $\epsilon_1$ which implies superplanckian field excursions and excludes the class of small field models. 

\subsection{Controversy about BICEP2 results}

It must be noted that BICEP2 results are still controversial.   Planck dust polarization maps have been recently published, and they suggest that the polarization fraction has considerable uncertainties, higher than the previous predictions that where used by BICEP2.   In the case there is no assumption about dust polarization, except the power spectrum shape
\be
C_l^{BB, dust} \propto l ^{-2.3}
\ee 
then it has been shown that solutions with no gravitational waves and $r < 0.11$ are favored at $2\sigma$ level. It is currently impossible to affirm with certainty that gravitational waves from inflation have been measured~\cite{Mortonson:2014bja,Flauger:2014qra,Ade:2014xna}.  One has to wait early 2015 for the Planck results on the CMB polarization. Planck is able to detect a tensor to scalar ratio $r \simeq 0.2$, but not $r\simeq 0.1$.  Therefore it is likely that we will have to wait for further observations, e.g. by BICEP3.  Note also that a joint more precise analysis between Planck and BICEP has been planned. 

\subsection{Future experiments}

There exists several projects of CMB experiments that could take place in the next 15 years.  Besides BICEP3, one can mention LiteBIRD~\cite{2014JLTP..176..733M}, the Cosmic Origin Explorer (COrE)~\cite{2011arXiv1102.2181T} and the Polarized Radiation Imaging and Spectroscopy (PRISM) missions~\cite{Andre:2013afa}, the latter two being now gathered in the COrE+ proposal.  One aim is to detect primordial gravitational waves at the $r \sim 10^{-3}$ level.   It is also possible to have access to a broader range of wavelength modes by measuring spectral distortions of the CMB black-body spectrum, e.g. with the Primordial Inflation Explorer (PIXIE)~\cite{2011JCAP...07..025K} or with PRISM.  Silk damped acoustic oscillations induce some energy injection in the CMB monopole, which results in spectral distortions that can therefore probe the scalar power spectrum on much smaller scales than with CMB temperature anisotropies.  A variety of inflation models could be tested with CMB distortions~\cite{Clesse:2014pna}

From 2020, additional constraints should come from large scale structure observations with Euclid, which is expected to improve by a factor 2-3 the present limits on the parameters describing the shape of the scalar power spectrum~\cite{2013LRR....16....6A}.  The 21cm signal from reionization and from the end of the dark ages could be also a powerful probe for cosmology (see e.g. Refs.~\cite{Mao:2008ug,Clesse:2012th,Brax:2012cr}), and especially for inflation.   Future giant radio-telescopes like the Square Kilometer Array (SKA) should inaugurate the detection of the 21cm signal, and an interesting concept of full digital kilometer size radio telescope dedicated to 21cm cosmology has been proposed~\cite{Tegmark:2008au}, the Fast Fourier Transform Telescope (FFTT).

\section{Model comparison:  a Bayesian approach} \label{sec:modelanalysis}

Considering only single field models, a plethora of scalar field potentials have been proposed, arising in various high energy frameworks like axions, non-minimally coupled Higgs field, supersymmetry, supergravity, Grand Unified Theories, extra-dimensions, string theory, brane cosmology, loop quantum gravity.   Most models proposed so far have been listed in Ref.~\cite{Martin:2013tda} and their compatibility with observations has been analyzed.  But in order to compare models and to hunt what is the best one given the data, one needs to use Bayesian statistical methods.  

In this section, we briefly review the basics of Bayesian inference and define the Bayes factor that gives the posterior odds of some model given a reference model.   Then, the results for single field models of inflation and Planck data are presented.  Even if they must be taken with caution, the results for a joint Planck and BICEP2 analysis are also discussed.  

This section is mostly based on the recent work by J. Martin, C. Ringeval, R. Trotta and V. Vennin~\cite{Martin:2013nzq,Martin:2014lra}, to which we invite the interested reader to refer for further details.  

\subsection{Notions of Bayesian inference}

Bayesian inference is based on the Bayes' theorem that determines how likely are some hypothesis given new data and some prior information.  To illustrate how works the theorem, let first consider a simple \textit{wikipedia} example:  suppose that a friend told you he met someone at a party.  If you don't have any further information and assuming that your friend does not talk more likely to women than men, then the probability that the person is a women is 50\%, which is denoted $P(W) = 0.5$ and $P(M) = 0.5$ ($M$ for man and $W$ for woman). Now if you understand in the conversation that this person had long hair, you may want to calculate the probability that the person was a woman.  If it is known that 75\% of women have long hair, which we denote $P(L|W)= 0.75$ (L for long hair), and that only 15\% of men have long hair, $P(L|M)= 0.15$ (those are conditional probabilities), then you can calculate the probability that a person is a women given that he or she is a long-haired person, $P(W|L)$.  It is obvious that  $P(W|L) P(L) = P(L|W) P(W) $, which can be rewritten 
\be
P(W|L) = \frac{P(L|W) P(W)} {  P(L|W) + P(L|M) } = \frac{0.75 \times 0.50}{0.75 \times 0.50 + 0.15 \times 0.50} = \frac{5}{6}
\ee
Now let us consider a model $M_i$ having continuous parameters $\theta_{ij}$, and that we want to evaluate the probability of some parameter values given some new data $D$.  The Bayes' theorem is generalized as follows:
\be
p(\theta_{ij}| D, \mathcal M_i ) = \frac{ \pi(\theta_{ij}| \mathcal M_i) \mathcal L(\theta_{ij}) }{\mathcal E (D| M_i )},
\ee
where $ \mathcal  L(\theta_{ij} )\equiv P(D| \theta_{ij}, \mathcal M_i)$ is the the so-called \textit{likelihood function} for the model parameters.  The function $\pi$ contains the prior information on the model parameters.  $\mathcal E$ is called the \textit{Bayesian evidence} and is defined as
\be
\mathcal E (D| M_i ) = \int \dd \theta_{ij} \mathcal L(\theta_{ij}) \pi(\theta_{ij } | \mathcal M_i ).
\ee
Note that it is just a normalization factor, and therefore it is not required to evaluate the Bayesian evidence if one just want to constrain the parameters $\theta_{ij}$.  On the other hand, if one needs the posterior probability of some model, one has to evaluate
\be
p(\mathcal M_i | D )= \frac{\pi(\mathcal M_i) \mathcal E(D | \mathcal M_i)}{\sum_i \pi(\mathcal M_i) \mathcal E(D | \mathcal M_i )  }
\ee
and then the Bayesian evidence needs to be computed.  In one consider that all single field models are known and that none of them is a priori favored, one has $\pi(\mathcal M_i) = 1/N_{\rr{models}}$.   The posterior odds of some model $\mathcal M_i$ compared to another reference model $\mathcal M_{\rr{Ref}}$ are encoded in the \textit{Bayes factor}
\be
B_{\rr{Ref}}^i \equiv \frac{\mathcal E(\mathcal M_{\rr{Ref}} | D )}{\mathcal E(\mathcal M_i | D )} =  \frac{p(\mathcal M_i | D )}{p(\mathcal M_{\rr{Ref}} | D )} .
\ee
It is interesting to note that the Bayes factor take into account the Occam's razor effect, in the sense that models with less parameters will be favored against models with more parameters and same predictions.  Another important point is that the Bayes factor depends on the prior given to the parameters $\theta_{ij}$.  Different priors change the model Bayesian evidence.  Thus for inflation, it is actually possible that a same scalar field potential arising in different frameworks, motivating different priors, has different Bayes factor depending on the considered scenario.  

The principal difficulty is to estimate the likelihood function $  \mathcal  L(\theta_{ij}) = P(D| \theta_{ij}, \mathcal M_i) $ in the full parameter space, for each model.  Usually, this is done by using Markov-Chain-Monte-Carlo (MCMC) methods.  Compared to standard Monte-Carlo that samples the parameter space with points randomly distributed, a Markov chain in which each point depends on the previous one is built and the statistical distribution of the points converges through the likelihood function.  The main advantage of MCMC methods is that the convergence is nearly linear, which allows to probe high-dimensional parameter spaces.  

\subsection{The best inflationary model after Planck}

\begin{figure}[ht] 
 	\begin{center}
	\includegraphics[height=100mm]{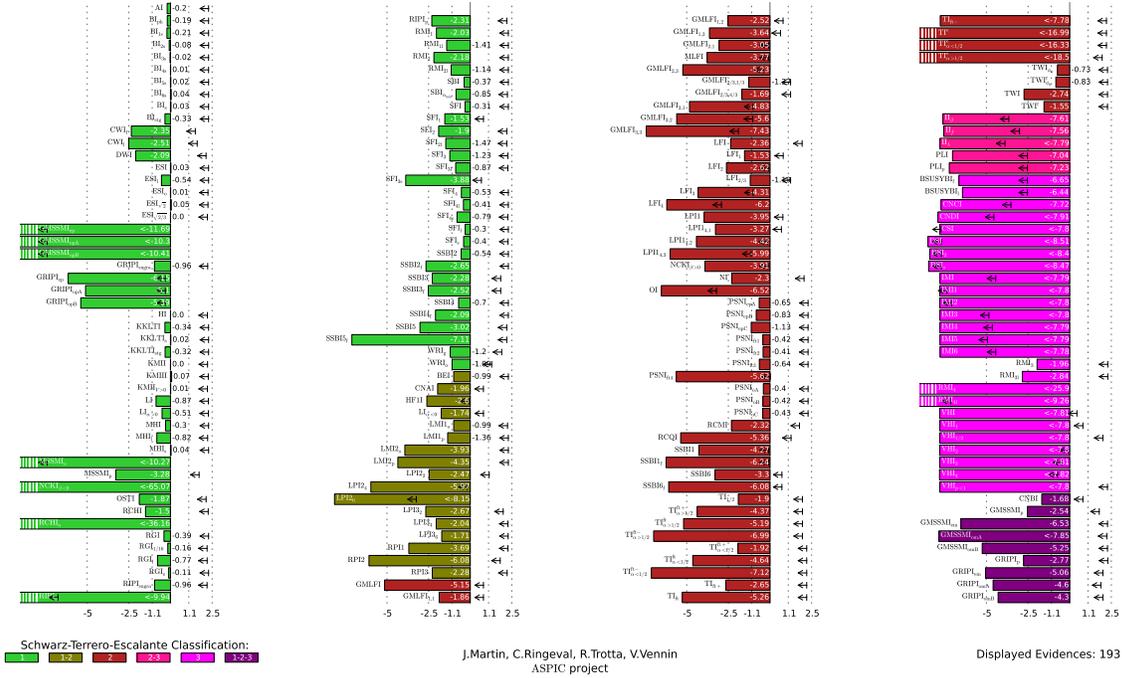}
	\caption{Figure fom~\cite{Martin:2013nzq} displaying the Bayes factors for all inflationary scenarios listed in the ASPIC library~\cite{Martin:2013tda}, using Higgs inflation as a reference model.  The bar color denote the class to which the model belongs (using Schwartz-Terrero-Escalante classification).  Arrows indicate the maximum likelihood}  
	 \label{fig:bestevid}
\end{center} 
\end{figure}

The Bayes actor has been calculated in~\cite{Martin:2013nzq} for Planck data and for all the 193 single field scenarios listed in Ref.~\cite{Martin:2013tda}, using the Higgs model (the only single field model with no free parameter) as a reference.  Those results are reported in Fig.~\ref{fig:bestevid}.  One can use the Jeffrey's scale for evaluating the Bayesian evidence between two models:  if $ |\ln B_{\rr{Ref}}^i | <1 $ the result is said to be \textit{inconclusive}; in the range  $1<  |\ln B_{\rr{Ref}}^i | <2.5$ there is a \textit{weak evidence}; for  $2.5<  |\ln B_{\rr{Ref}}^i | <5$ there is a \textit{moderate evidence}; if  $|\ln B_{\rr{Ref}}^i | >5$ there is a $\textit{strong evidence}$.  

The first important result is that Planck data rule out at a strong evidence level about one third of the models.  This demonstrates that Planck results are very impressive.  Only one fourth of models are found to be inconclusive compared to the best model.  Interestingly, all the favored scenarios have a scalar field potential of plateau-type.  Among them, Higgs inflation (or equivalently the Starobinsky model) is at the top, whereas it is totally predictive since it has no free parameter.  

In Higgs inflation, the inflaton field is identified to the Higgs field $h$ recently discovered at the Large Hadron Collider.  But for being in agreement with data, the model requires that the Higgs is non-minimally coupled to gravity.   This is the simplest model arising in the context of the standard model of particle physics.   In the Einstein frame, the scalar field potential is of the form
\be
V(\phi) = \Lambda^4 \left( 1 - \rr e^{- \sqrt{2/3} \phi/ \Mpl} \right) ^2,
\ee
and the only parameter $\Lambda$ is fixed by normalizing the scalar power spectrum amplitude to Planck measurement, so that there no remaining free parameter.  The Higgs model is equivalent to the Starobinsky model, based on the action $S = \frac 1 2 \int \dd x^4 \sqrt{-g} ( \Mpl^2 R + R^2/ 6M^2) $.

\subsection{The best inflationary model after Planck+BICEP2}

A similar analysis has been performed for Planck plus BICEP2~\cite{Martin:2014lra}, assuming that the B-modes detected by BICEP2 are of primordial origin, and using only the first four band powers (the others preferring a large value of the tensor-to-scalar ratio already strongly disfavored by Planck and WMAP).   

\begin{figure}[ht] 
 	\begin{center}
	\includegraphics[height=150mm]{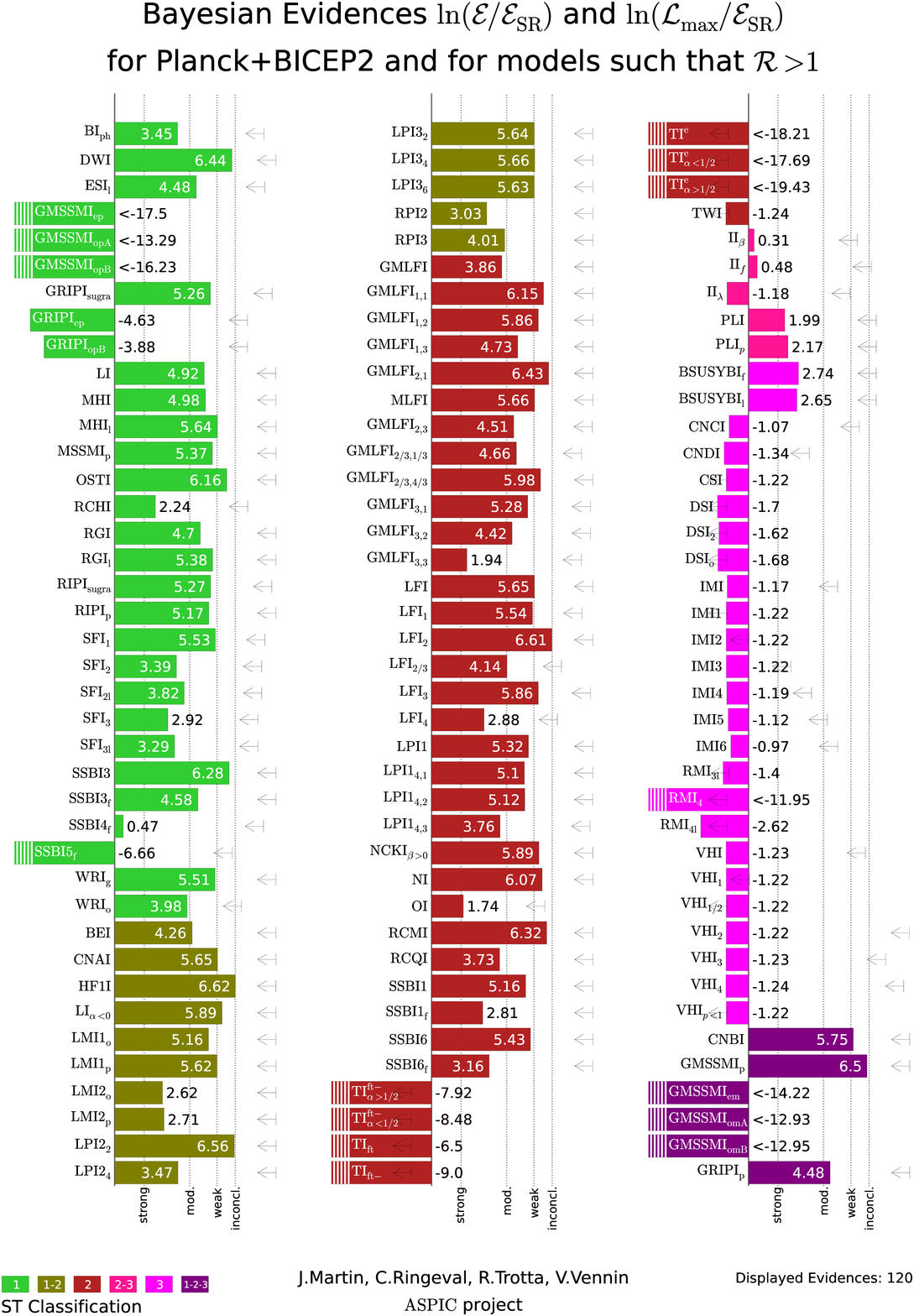}
	\caption{Figure fom~\cite{Martin:2014lra} displaying the Bayes factors for all inflationary scenarios listed in the ASPIC library~\cite{Martin:2013tda} compatible both with Planck and BICEP2, using a slow-roll model as a reference.  As in the previous figure, the bar color denote the class to which the model belongs and arrows indicate the maximal likelihood value}  
	 \label{fig:bestevid2}
\end{center} 
\end{figure}

The principal finding is that there exist a net tension between Planck and BICEP2 data, most models favored by Planck alone being disfavored by BICEP2 alone (in terms of Bayesian evidence), and inversely.   There nevertheless remains a series of models that are compatible both with Planck and BICEP2, and whose Bayes factor is reported in Fig.~\ref{fig:bestevid2}.  Note that the reference model is not Higgs inflation (which is not compatible with BICEP2) but a slow-roll model with $\epsilon_1$, $\epsilon_2$ and $\epsilon_3$ as free parameters.  In this case, it is found that large field models are favored and the simple potential $V(\phi ) = m ^2 \phi^2$ is one of the best models. 

\section{Open problems}  \label{sec:openquest}

In this section I briefly mention a discuss a few open issues related to inflation.

\subsection{Inflation vs. alternatives:}   Alternative to inflation exist and include string gas cosmology, matter bounces, ekpyrotic/cyclic scenarios (for a review, see e.g. Rev.~\cite{Brandenberger:2009jq}).  But most of them do not solve simultaneously all the problems of the standard cosmological scenario, and thus the inflationary paradigm is often considered as the best option, even if it must be seen as a model among others and in no way a theory already proved by observations.  Even the detection of B-modes should not be considered as a proof of inflation.  Nevertheless, single field models of inflation have specific predictions such as the consistency relation between the scalar and tensor spectral indexes.  Checking observationally that it is verified would strongly disfavor alternatives to inflation, and thus this will be one major objective of future CMB experiments. 

\subsection{Reheating:}  

At the end of inflation, the energy stored in scalar field potential must decay into standard model particles.   It is thought that when inflation stops, the field start to oscillate around the minimum of the potential.  The coherent oscillations can be considered as a collection of independent scalar particles.  If they couple to other particles, the inflaton can decay perturbatively to produce light particles.  
Another possibility is a phase of tachyonic preheating, which occurs in hybrid inflation scenarios due to the exponential grow of the modes of an auxiliary field during the waterfall phase.  The modes become quickly non-linear and reheating occurs with the dissipation of interacting classical field oscillations 

As we have shown, the observable predictions for a single field model in slow-roll are easily calculated when one knows $t_*$, the time at which the observable pivot scale $k_*$ exit the Hubble radius during inflation.   But to find $t_*$, one does not only needs to solve the dynamics of inflation but also all the subsequent expansion history.  The simplest assumption is to consider that the reheating is instantaneous and that a radiation era is immediately triggered at the end of inflation.  But if this is not the case, then the reheating history affects the observable predictions for a given inflation model.  One has therefore to derive reheating consistent constraints on its parameter space, e.g. by parametrizing the reheating phase by its duration and its mean equation of state (see e.g.~\cite{Martin:2014nya}).

\subsection{Initial homogeneity problem:}  

The question of how homogeneous must have been the initial conditions for inflation to be triggered has been tackled for more than twenty years by several authors~\cite{Goldwirth:1989pr,Goldwirth:1989vz,Goldwirth:1991rj,KurkiSuonio:1993fg,Laguna:1991zs,Deruelle:1994pa}.  It has been shown for the simplest scalar field potentials that the Universe must have been initially homogeneous on scales larger than the Hubble radius.  To draw this conclusion, relativistic simulations of the pre-inflation era have been conducted, either by using gradient expansion methods or by solving the full Einstein equations, in 1+1 dimensions (spherically symmetric case) or in 3+1 dimensions.
Thus inflation merely transforms one problem of homogeneity into another one, since an incredibly homogeneous initial state is required.   This issue is called the \textit{Homogeneity Problem} for inflation.  

\subsection{Trans-Planckian problem:}  

In inflation observable scales exit the Hubble radius about 60 e-folds before the end of inflation.   But if inflation lasted for much more than 60 e-folds, then the observable wavelength modes were initially not only sub-Hubble but also sub-Planckian, in a regime where the physical principles which underlie the calculations of the power
spectrum are possibly not valid anymore.   This occurs whereas the energy density itself is well below the Planck scale and so the homogeneous classical field dynamics is perfectly valid.  There are several approaches to model trans-planckian effects, e.g. by modifying the dispersion relation, $k \rightarrow w(k)$.  One of them is to assume 
that Fourier modes are created 
when their wavelength equals some critical scale denoted by $M_c$.   This modification of the infrared behavior of the perturbation modes induce oscillations in the scalar power spectrum~\cite{Martin:2000xs,Martin:2004iv}  
\be
\mathcal P_\zeta (k) = \mathcal P_\zeta^{\rr{std}} \times \left\{ 1- 2  |x| \sigma_0 \cos \left[  \frac{2 \epsilon_1}{\sigma_0} \ln \left( \frac{k}{\kp} \right)+ \psi \right]   \right\}
\ee
where $\sigma_0 \equiv H/M_c$,  $\epsilon_1, \epsilon_2 $ are the usual Hubble-flow parameters.   $x$ is a complex number of modulus $ |x|$ and phase $\phi$ that parameterizes the initial conditions for the modes.   Constraining the presence of oscillatory features in the scalar power spectrum therefore also constraints trans-Planckian physics.  


\subsection{Avoiding the Big-Bang singularity}

Models of inflation alone are insufficient to describe the initial state of the Universe. 
They do not solve the initial singularity problem so that a quantum theory of gravitation is needed to describe the Universe at the Planck energy scale.   
But it is possible that the Universe was dominated by curvature just before inflation is triggered.  In this case general relativity allows that the Universe has performed a classical bounce, thus avoiding the initial singularity, followed by a phase of inflation to explain the apparent flatness today.  This scenario leads to specific signatures in the scalar power spectrum, taking the form of superimposed oscillations~\cite{Lilley:2011ag}.  Compared to the case of trans-Planckian effects, there are two important differences:  for trans-planckian effects, the oscillatory term has a $\ln k$ dependance rather than a linear $k$ dependance for bounding cosmologies.  

\subsection{Eternal inflation:}  

In most inflation models, there exist regions in the field space where the potential is so flat that its quantum fluctuations $\Delta \phi^{\rr{qu}} \sim H/2 \pi$ in a Hubble time dominate over its classical evolution $\Delta \phi^{\rr{cl}} = \dot \phi / H$.  Regions in the real space where quantum fluctuations push the field through values where the potential is more flat expand faster than others, and they can experience in turn quantum fluctuations pushing the field towards more flat regions of the potential, which expand faster, and so on, so that they occupy a never-ending increasing part of the total volume.  Inflation globally never ends, this is called the eternally self-reproducing regime.  In this picture, our observable patch of the Universe emerges from an eternally inflating multiverse.  The problem is that  most models contains a self-reproducing regime.  
But eternal inflation leads to a collection of difficult problems like the measure non-normalisability and the apparent lost of unitarity (see e.g Refs.~\cite{Linde:2010xz,Guth:2011ie} for a review on eternal inflation and related issues). 

\section*{Acknowledgments}

It is a pleasure to thank the organizers of the X Modave School of Mathematical Physics for their invitation, especially Ruben Monten, as well as other lecturers and all the participants for interesting and stimulating discussion.   My work is supported by the \textit{Return grant} program of the Belgian Federal Science Policy (BELSPO).  

\bibliographystyle{hunsrt}
\bibliography{biblio}

\end{document}